\documentstyle[prb,aps]{revtex}

\begin{document}
\draft
\title{%
  Many-body theory of pump-probe spectra for highly excited
  semiconductors}
\author{T.~J.~Inagaki$\mbox{}^{\ast}$ \cite{E-mail},
        T.~Iida$\mbox{}^{\ddag}$ and
        M.~Aihara$\mbox{}^{\ast}$}
\address{
  $\mbox{}^{\ast}$
  Graduate School of Materials Science,
  Nara Institute of Science and Technology,
  Ikoma, Nara 630-0101, Japan   \\
  $\mbox{}^{\ddag}$
  Department of Physics, Osaka City University, 
  Sumiyoshi-ku, Osaka 558-8585, Japan}
\date{\today}

\maketitle

\begin{abstract}
  We present a unified theory for pump-probe spectra in highly excited
  semiconductors, which is applicable throughout the whole density
  regime including the high-density electron-hole BCS state and the
  low-density excitonic Bose-Einstein condensate (BEC).
  The analysis is based on the BCS-like pairing theory combined with
  the Bethe-Salpeter (BS) equation, which first enables us to
  incorporate the state-filling effect, the band-gap renormalization
  and the strong/weak electron-hole pair correlations in a unified
  manner.
  We show that the electron-hole BCS state is distinctly stabilized
  by the intense pump-light, and this result strongly suggests that
  the macroscopic quantum state can be observed under the strong
  photoexcitation.
  The calculated spectra considerably deviate from results given by
  the BCS-like mean field theory and the simple BS equation without
  electron-hole pair correlation especially in the intermediate
  density states between the electron-hole BCS state and the excitonic
  BEC state. 
  In particular, we find the sharp stimulated emission and absorption
  lines which originate from the optical transition accompanied by the 
  collective phase fluctuation mode in the electron-hole BCS state.
  From the pump-probe spectral viewpoint, we show that this
  fluctuation mode changes to the exciton mode with decreasing
  carrier density.
\end{abstract}


\section{Introduction}
\label{sec:1}

The many-body effects of electron--hole systems in semiconductors
generated by the intense laser field have long been attractive
subjects \cite{Haug,Hanamura}.
One of the most outstanding characteristics of the system is that the 
electron-hole density and the macroscopic quantum state can easily be
controlled by changing the frequency and the intensity of excitation
light.
Recent developments in the experimental techniques make it possible
to observe the remarkable phenomena suggesting the optically induced
macroscopic quantum states.
In particular, the anomalous exciton transport observed in
Cu$_2$O (Ref.~\onlinecite{Fortin}) and BiI$_3$
(Ref.~\onlinecite{Karasawa})  are the typical examples of them. 
These phenomena become more significant with increasing the
electron-hole densities, and are in a marked contrast to the simple
ballistic exciton propagation or the conventional diffusive exciton 
transport.

In order to understand the nature of the macroscopic quantum state of
electron-hole systems, conventional approaches based either on the
weak interacting Boson model \cite{Hanamura} or on the BCS-like
mean-field theory \cite{Comte} are inappropriate. 
This is because the mean interparticle distance in the highly
photoexcited system is often of the same order as the radius of the
bound electron-hole pairs, and the system is in the intermediate
state between the excitonic Bose-Einstein condensate (BEC) and the
electron-hole BCS state.
The nature of the system is characterized by two kinds of order: the
$\mbox{}^4$He-like order associated with the center-of mass motion of
electron-hole pairs and the BCS-like order associated with the
electron-hole pair formation. 
In order to investigate the physical properties of the system, we have
to incorporate the state-filling effect, which reflects the Pauli
principle, and the center-of-mass motion of electron-hole pairs on the
same footing.

In our previous work\cite{inagaki}, we have presented a theory of
luminescence spectra for electron-hole systems in highly excited
semiconductors; this theory gives the correct description of
the luminescence spectra in the crossover regime between the
low-density excitonic BEC states and the high-density electron-hole
BCS states.
The analysis is based on the generalized random-phase approximation
\cite{Anderson} combined with the Bethe-Salpeter (BS) equation
\cite{Nakanishi}.
This approach enables us to incorporate the strong electron-hole pair
correlation and the quantum fluctuation effect associated with the
center-of-mass motion of electron-hole pairs. 
We have shown that the crossover from the excitonic BEC state to the 
electron-hole BCS state manifests itself in the luminescence spectra.
In particular, it is found that the broad emission band, arising from
the pair recombination in the electron-hole BCS state, splits into the
P and P$\mbox{}_2$ emission bands \cite{P-line,Kawasaki} with
decreasing the electron-hole density.
These results are closely related to the recent noteworthy experiment
for ZnO, in which the room temperature ultraviolet laser emission is
observed \cite{Kawasaki2}.

The purpose of the present work is to clarify the properties of the 
high-density electron-hole systems subject to the intense light by
analyzing the pump-probe spectrum.
To our knowledge, a many-body theory for absorption spectrum of
semiconductors was first presented by L\"owenau 
{\it et al.}\cite{LSH}; their analysis can be applied both to the low-
and high-density states.
They solved the BS equation for the interband dielectric function by a
numerical matrix inversion \cite{SLH,HK}, and reproduces a series of
excitonic sharp spectral components in the low particle densities.
However, the electron-hole pair correlation in the high-density regime
is not taken into account, so that the many exciton effect and the
BCS-like macroscopic quantum state cannot be incorporated.

In Ref.~\onlinecite{Comte2}, the effect of pair correlation is taken
into account for the pump-probe spectrum by the simple BCS-like
mean-field theory at zero temperature.
By solving the BCS-like gap equation for electron-hole systems, they
found that the electron-hole pair correlations are considerably
enhanced by the pump-light intensity (``light enhanced excitons'').
Their calculated spectra exhibit the transparent region between the
stimulated emission and the absorption components, which is the
manifestation of the BCS-like gap formation.
Whereas their theory takes into account the pair correlation in a
self-consistent manner, they neglected the screening effct of the
Coulomb interaction and the multiple Coulomb interaction between
electron-hole pairs. 
As a result, the excitonic sharp spectral components do not reproduce
in the low electron-hole densities.

The discussion was further extended to take into account the screening
effect of the Coulomb interaction and the finite temperature effect by
Iida {\it et al.}  \cite{Iida}
They considered the screening effect by the quasi-static RPA due to
Haug and Schmitt-Rink \cite{Haug}, and found that the bistable
behavior arises from the band-gap renormalization.
However, their analysis is based on the simple BCS-like mean field
theory, so it cannot properly describe the excitonic structure in the
low particle densities.

In the present article, we give a unified theory of the pump-probe
spectra for highly excited semiconductors, which can be applied
throughout the whole density range including the high-density
electron-hole BCS state and the low-density excitonic BEC states. 
We consistently solve the combined equations of the BCS-like gap
equation and the BS equation for retarded electron-hole pair Green
function, which first enables us to consider the state-filling
effect, the band-gap renormalization, and the weak/strong pair
correlation in a unified manner.
This analysis is closely related to that in Ref.~\onlinecite{Chu},
where the absorption-gain spectra for condensed exciton systems are
calculated for various electron-hole densities and temperatures by
properly taking into account the ladder diagram for vertex function.
In the present article, we calculate the pump-probe spectra for
various excitation conditions and systematically analyze the
dependence of the spectra on the pump-light intensity and on the
pump-light frequency.

As a result, we find that the electron-hole BCS state is noticeably
enhanced by the intense pump-light; this result strongly suggests that
the macroscopic quantum state can be observed under the strong
photoexcitations. 
Moreover, in the high-density state generated by the small
off-resonant excitation, the calculated spectrum shows a pair of sharp
peaked structures at the edge of the broad absorption and stimulated
emission bands.
In contrast to this, we find a series of sharp exciton lines in the
low density regime generated by the large off-resonant excitation.
In the intermediate density generated by the strong photoexcitation,
the calculated spectra considerably deviate from results obtained by
the BCS-like mean-field analysis and those by the simple BS equation
without pair correlation.
In particular, we find the sharp spectral components originating from
the excitation of the collective phase fluctuation mode of
electron-hole BCS state (similar to the Anderson mode in
superconductivity), which change to the exciton series with 
decreasing carrier density.

The article is organized as follows.
In Sec.~\ref{sec:2}, we derive the BS equation for the retarded
Green function of electron-hole pairs 
within the quasi-statically screened ladder approximation.
In order to incorporate the pairing effect to the pump-probe spectra, 
we express the electron-hole pair Green function with respect to the
field operators of Bogolons, that are the elementary excitation of 
electron-hole BCS state.
We numerically calculate the pump-probe spectra by solving the BS
equation in Sec.~\ref{sec:3}. 
Finally, discussions and conclusion are given in Sec.~\ref{sec:4}.

\section{Formulation}
\label{sec:2}
\subsection{Model Hamiltonian}
We consider an electron-hole system in a direct-gap semiconductor,
which consists of the isotropic, nondegenerate parabolic conduction
and valence bands.
The repulsive interactions between electrons and between holes as well
as the electron-hole attractive interaction are taken into account.
The spin degrees of freedom are neglected to avoid the unnecessary
complication due to the biexciton state.
We consider that the system is in a stationary state generated
by an intense monochromatic pump light with frequency 
$ \omega_{\rm L} $; the pump light is treated as a classical field while
the probe light is quantized.
The model Hamiltonian is expressed as 
\begin{mathletters}
\begin{equation}
  \label{eq:Hamiltonian}
  H_{\rm tot} = H_{\rm r} + H_{\rm eh} + H_{\rm pump} + H_{\rm probe},
\end{equation}
where  $ H_{\rm r} $, $ H_{\rm eh} $, $ H_{\rm pump} $ and 
$ H_{\rm probe} $ are the Hamiltonian of the probe photon,
the electron-hole Hamiltonian, the interaction Hamiltonian between
particles and pump light, and the interaction Hamiltonian between
particles and probe light, respectively. 
These Hamiltonians are expressed in terms of the field operators of
electrons ($ c_{\mbox{\boldmath \scriptsize $p$}} $), holes
($ d_{\mbox{\boldmath \scriptsize $p$}} $) and photons ($ b $) as
follows, 
\begin{eqnarray}
  \label{eq:2-1}
  H_{\rm r} & = & \omega b^{\dagger} b ,
\nonumber \\
  H_{\rm eh} & = &
     \sum_{\mbox{\boldmath \scriptsize $k$}} \left\{
       \varepsilon^{e}_{\mbox{\boldmath \scriptsize $k$}}
       c^{\dagger}_{\mbox{\boldmath \scriptsize $k$}}
       c_{\mbox{\boldmath \scriptsize $k$}}
     + \varepsilon^{h}_{\mbox{\boldmath \scriptsize $k$}}
       d^{\dagger}_{-\mbox{\boldmath \scriptsize $k$}}
       d_{-\mbox{\boldmath \scriptsize $k$}}
     \right\}
\nonumber  \\
  && +
    \frac{1}{2}
    \sum_{\mbox{\boldmath \scriptsize $k$},
          \mbox{\boldmath \scriptsize $p$},
          \mbox{\boldmath \scriptsize $q$}}
    V_{\mbox{\boldmath \scriptsize $q$}}
    \left\{
      c_{\mbox{\boldmath \scriptsize $k$}
        +\mbox{\boldmath \scriptsize $q$}}^{\dagger}
      c_{\mbox{\boldmath \scriptsize $p$}
        -\mbox{\boldmath \scriptsize $q$}}^{\dagger}
      c_{\mbox{\boldmath \scriptsize $p$}}
      c_{\mbox{\boldmath \scriptsize $k$}}
   +
      d_{\mbox{\boldmath \scriptsize $k$}
        +\mbox{\boldmath \scriptsize $q$}}^{\dagger}
      d_{\mbox{\boldmath \scriptsize $p$}
        -\mbox{\boldmath \scriptsize $q$}}^{\dagger}
      d_{\mbox{\boldmath \scriptsize $p$}}
      d_{\mbox{\boldmath \scriptsize $k$}}
    -
      2
      c_{\mbox{\boldmath \scriptsize $k$}
        +\mbox{\boldmath \scriptsize $q$}}^{\dagger}
      c_{\mbox{\boldmath \scriptsize $k$}}
      d_{\mbox{\boldmath \scriptsize $p$}
        -\mbox{\boldmath \scriptsize $q$}}^{\dagger}
      d_{\mbox{\boldmath \scriptsize $p$}}
    \right\} ,
\nonumber \\
   H_{\rm pump} & = &
     \lambda
     \sum_{\mbox{\boldmath \scriptsize $k$}} 
     \left\{
       c_{\mbox{\boldmath \scriptsize $k$}}
       d_{-\mbox{\boldmath \scriptsize $k$}}
       \exp(i \omega_{\rm L} t)
     + d_{-\mbox{\boldmath \scriptsize $k$}}^{\dagger}
       c_{\mbox{\boldmath \scriptsize $k$}}^{\dagger}
       \exp(-i \omega_{\rm L} t)
    \right\} ,
\nonumber \\
   H_{\rm probe} & = &
     \sum_{\mbox{\boldmath \scriptsize $k$}} 
     \left\{
       g_{\mbox{\scriptsize \boldmath $k$}} 
       b^{\dagger}
       c_{\mbox{\boldmath \scriptsize $k$}}
       d_{-\mbox{\boldmath \scriptsize $k$}}
     + g_{\mbox{\scriptsize \boldmath $k$}} ^{\ast}
       d_{-\mbox{\boldmath \scriptsize $k$}}^{\dagger}
       c_{\mbox{\boldmath \scriptsize $k$}}^{\dagger} b
    \right\} ,
\end{eqnarray}
\end{mathletters}
where $ \omega $ is the frequency of probe photon.
Here $ g_{\mbox{\boldmath \scriptsize $k$}} $ is the coupling
constant between particles and the probe light;
$ \lambda $ is the interaction energy between particles and
the pump light. 
The single-particle energies of electrons and holes are given by
$ \varepsilon^{e}_{\mbox{\boldmath \scriptsize $k$}} =
  k^2/(2 m_e) + E_g $, and
$ \varepsilon^{h}_{\mbox{\boldmath \scriptsize $k$}} = k^2/(2 m_h) $,
respectively, where $ m_e $ ($m_h$) and $ E_g $ are the electron
(hole) effective mass and band-gap energy, respectively.
The Coulomb interaction is written as 
$ V_{\mbox{\boldmath \scriptsize $q$}} = 4 \pi e^2/(\epsilon_0 q^2) $,
where  $ \epsilon_0 $ is the background dielectric constant of the
unexcited crystal.

When we analyze a matter which is driven by a coherent light, it is
convenient to convert the matter Hamiltonian to the time independent
form in the so-called rotating frame.
For this purpose, we use the unitary transformation given by the
unitary operator \cite{Galitskii},
\begin{eqnarray}
  \label{eq:2-2a}
  T(t) = \exp \left[
    - \frac{i \omega_{\rm L} t}{2}
    \sum_{\mbox{\boldmath \scriptsize $k$}}
    \left(
      c_{\mbox{\boldmath \scriptsize $k$}}^\dagger
      c_{\mbox{\boldmath \scriptsize $k$}}
    + d_{-\mbox{\boldmath \scriptsize $k$}}^\dagger
      d_{-\mbox{\boldmath \scriptsize $k$}}
    \right)
  \right] .
\end{eqnarray}
The transformed Hamiltonian is written as follows,
\begin{eqnarray}
  \label{eq:2-2b}
  \tilde{H}_{\rm eh} & = &
     \sum_{\mbox{\boldmath \scriptsize $k$}} \left\{
       \xi^{e}_{\mbox{\boldmath \scriptsize $k$}}
       c^{\dagger}_{\mbox{\boldmath \scriptsize $k$}}
       c_{\mbox{\boldmath \scriptsize $k$}}
     + \xi^{h}_{\mbox{\boldmath \scriptsize $k$}}
       d^{\dagger}_{-\mbox{\boldmath \scriptsize $k$}}
       d_{-\mbox{\boldmath \scriptsize $k$}}
     \right\}
\nonumber  \\
  && +
    \frac{1}{2}
    \sum_{\mbox{\boldmath \scriptsize $k$},
          \mbox{\boldmath \scriptsize $p$},
          \mbox{\boldmath \scriptsize $q$}}
    V_{\mbox{\boldmath \scriptsize $q$}}
    \left\{
      c_{\mbox{\boldmath \scriptsize $k$}
        +\mbox{\boldmath \scriptsize $q$}}^{\dagger}
      c_{\mbox{\boldmath \scriptsize $p$}
        -\mbox{\boldmath \scriptsize $q$}}^{\dagger}
      c_{\mbox{\boldmath \scriptsize $p$}}
      c_{\mbox{\boldmath \scriptsize $k$}}
   +
      d_{\mbox{\boldmath \scriptsize $k$}
        +\mbox{\boldmath \scriptsize $q$}}^{\dagger}
      d_{\mbox{\boldmath \scriptsize $p$}
        -\mbox{\boldmath \scriptsize $q$}}^{\dagger}
      d_{\mbox{\boldmath \scriptsize $p$}}
      d_{\mbox{\boldmath \scriptsize $k$}}
    -
      2
      c_{\mbox{\boldmath \scriptsize $k$}
        +\mbox{\boldmath \scriptsize $q$}}^{\dagger}
      c_{\mbox{\boldmath \scriptsize $k$}}
      d_{\mbox{\boldmath \scriptsize $p$}
        -\mbox{\boldmath \scriptsize $q$}}^{\dagger}
      d_{\mbox{\boldmath \scriptsize $p$}}
    \right\} ,
\nonumber \\
   \tilde{H}_{\rm pump} & = &
     \lambda
     \sum_{\mbox{\boldmath \scriptsize $k$}} 
     \left\{
       c_{\mbox{\boldmath \scriptsize $k$}}
       d_{-\mbox{\boldmath \scriptsize $k$}}
     + d_{-\mbox{\boldmath \scriptsize $k$}}^{\dagger}
       c_{\mbox{\boldmath \scriptsize $k$}}^{\dagger}
    \right\} ,
\nonumber \\
   \tilde{H}_{\rm probe} & = &
     \sum_{\mbox{\boldmath \scriptsize $k$}}
     \left\{
       g_{\mbox{\scriptsize \boldmath $k$}} 
       b^{\dagger}
       c_{\mbox{\boldmath \scriptsize $k$}}
       d_{-\mbox{\boldmath \scriptsize $k$}}
       \exp(-i \omega_{\rm L} t)
     + g_{\mbox{\scriptsize \boldmath $k$}}^{\ast}
       d_{-\mbox{\boldmath \scriptsize $k$}}^{\dagger}
       c_{\mbox{\boldmath \scriptsize $k$}}^{\dagger} b
       \exp(i \omega_{\rm L} t)
    \right\} ,
\end{eqnarray}
where 
$ \xi^{e,h}_{\mbox{\boldmath \scriptsize $k$}} =
  \varepsilon^{e,h}_{\mbox{\boldmath \scriptsize $k$}}
  - \omega_{\rm L}/2 $.
The resulting stationary state of driven electron-hole system is
controlled by the frequency of pump light $ \omega_{\rm L} $, and the
interaction energy $ \lambda $ between pump light and carriers.
The Hamiltonian $ \tilde{H}_{\rm eh} $ is formally regarded as a
grand-canonical Hamiltonian of the electron-hole system, in which the
chemical potential of electron-hole pairs is given by 
$ \mu = \omega_L - E_g $.

\subsection{Optical response of probe light}
In order to evaluate the optical spectra, let us consider the
equation-of-motion for expectation value of the number operator for a
probe photon, $ N(t) = b^{\dagger}(t) b(t) $.
A perturbative calculation with respect to $ \tilde{H}_{\rm probe} $
gives
\begin{eqnarray}
  \label{eq:1-1a}
  \frac{d}{dt} \langle N(t) \rangle =
    \Pi^{<}(\omega) \langle N(t) + 1 \rangle
  - \Pi^{>}(\omega) \langle N(t) \rangle 
\end{eqnarray}
where $ \langle \cdots \rangle $ indicates the expectation value with
respect to the ground state of 
$ H_{\rm tot} $.
The quantities $ \Pi^{<}(\omega) $ and $ \Pi^{>}(\omega) $ are the
emission and the absorption rate of probe photon, respectively.
The pump-probe spectrum is then given by 
$ \Pi^{>}(\omega) - \Pi^{<}(\omega) $. 
The quantities $ \Pi^{<}(\omega) $ and $ \Pi^{>}(\omega) $ are
expressed in terms of electron and hole operators as following, 
\begin{eqnarray}
  \label{eq:1-1b}
  \Pi^{>}(\omega) & = &
    \sum_{\mbox{\boldmath \scriptsize $k$},
          \mbox{\boldmath \scriptsize $p$}}
    g_{\mbox{\boldmath \scriptsize $k$}}^{\ast}
    g_{\mbox{\boldmath \scriptsize $p$}}
    \int_0^\infty dt
      \langle
        d_{-\mbox{\boldmath \scriptsize $p$}}(t)
        c_{\mbox{\boldmath \scriptsize $p$}}(t)
        c_{\mbox{\boldmath \scriptsize $k$}}^{\dagger}(0)
        d_{-\mbox{\boldmath \scriptsize $k$}}^{\dagger}(0)
      \rangle_0
      {\rm e}^{i(\omega - \omega_{\rm L})t}
      + c.c.
\nonumber  \\
  & = & - 2 {\rm Im} G^{>}(\omega - \omega_{\rm L} + i \gamma) ,
\nonumber  \\
  \Pi^{<}(\omega) & = &
    \sum_{\mbox{\boldmath \scriptsize $k$},
          \mbox{\boldmath \scriptsize $p$}}
    g_{\mbox{\boldmath \scriptsize $k$}}^{\ast}
    g_{\mbox{\boldmath \scriptsize $p$}}
    \int_0^\infty dt
      \langle
        c_{\mbox{\boldmath \scriptsize $k$}}^{\dagger}(0)
        d_{-\mbox{\boldmath \scriptsize $k$}}^{\dagger}(0)
        d_{-\mbox{\boldmath \scriptsize $p$}}(t)
        c_{\mbox{\boldmath \scriptsize $p$}}(t)
      \rangle_0
      {\rm e}^{i(\omega - \omega_{\rm L})t}
      + c.c.
\nonumber  \\
  & = & - 2 {\rm Im} G^{<}(\omega - \omega_{\rm L} + i \gamma) ,
\end{eqnarray}
where $ \gamma $ is the exciton decay constant and 
$ \langle \cdots \rangle_0 $ stands for the expectation value with
respect to the ground state of 
$ H = \tilde{H}_{\rm eh} + \tilde{H}_{\rm pump} $.
Here $ G^{>}(\omega) $ and $ G^{<}(\omega) $ are the Fourier transform
of the correlation functions defined by
\begin{eqnarray}
  \label{eq:1-1c}
  G^{>}(t) & = & -i \Theta(t)
  \sum_{\mbox{\boldmath \scriptsize $k$},
        \mbox{\boldmath \scriptsize $p$}}
    g_{\mbox{\boldmath \scriptsize $k$}}^{\ast}
    g_{\mbox{\boldmath \scriptsize $p$}}
      \langle
        d_{-\mbox{\boldmath \scriptsize $p$}}(t)
        c_{\mbox{\boldmath \scriptsize $p$}}(t)
        c_{\mbox{\boldmath \scriptsize $k$}}^{\dagger}(0)
        d_{-\mbox{\boldmath \scriptsize $k$}}^{\dagger}(0)
      \rangle_0 ,
\nonumber  \\
  G^{<}(t) & = & -i \Theta(t)
  \sum_{\mbox{\boldmath \scriptsize $k$},
        \mbox{\boldmath \scriptsize $p$}}
    g_{\mbox{\boldmath \scriptsize $k$}}^{\ast}
    g_{\mbox{\boldmath \scriptsize $p$}}
      \langle
        c_{\mbox{\boldmath \scriptsize $k$}}^{\dagger}(0)
        d_{-\mbox{\boldmath \scriptsize $k$}}^{\dagger}(0)
        d_{-\mbox{\boldmath \scriptsize $p$}}(t)
        c_{\mbox{\boldmath \scriptsize $p$}}(t)
      \rangle_0 .
\end{eqnarray}
The pump-probe spectrum is given by
\begin{equation}
  \label{eq:pp}
  A(\omega) = -2 {\rm Im}G_{\rm R}(\omega - \omega_{\rm L} + i \gamma) ,
\end{equation}
where $ G_{\rm R}(\omega) $ is the Fourier transform of the retarded
Green function for electron-hole pairs given by
\begin{equation}
  \label{eq:rGF}
  G_{\rm R}(t) \equiv G^{>}(t) - G^{<}(t) .
\end{equation}
In the following analysis, we calculate $ G_{\rm R}(t) $ by solving
the BS equation.

\subsection{BCS-like gap equation for electron-hole systems}
In order to incorporate the electron-hole pair correlation, let
us introduce the Bogoliubov transformation given by, 
\begin{eqnarray}
\label{eq:BT}
  c_{\mbox{\boldmath \scriptsize $k$}} = 
  u_{\mbox{\boldmath \scriptsize $k$}}
  \alpha_{\mbox{\boldmath \scriptsize $k$}} +
  v_{\mbox{\boldmath \scriptsize $k$}} 
  \beta_{-\mbox{\boldmath \scriptsize $k$}}^{\dagger} ,
\nonumber \\
  d_{-\mbox{\boldmath \scriptsize $k$}} = 
  u_{\mbox{\boldmath \scriptsize $k$}}
  \beta_{-\mbox{\boldmath \scriptsize $k$}} -
  v_{\mbox{\boldmath \scriptsize $k$}} 
  \alpha_{\mbox{\boldmath \scriptsize $k$}}^{\dagger},
\end{eqnarray}
where 
$ \alpha_{\mbox{\boldmath \scriptsize $k$}} $ and
$ \beta_{-\mbox{\boldmath \scriptsize $k$}} $ are the annihilation
operators for new quasiparticles (Bogolons).
The Bogoliubov parameters
$ u_{\mbox{\boldmath \scriptsize $k$}} $ and 
$ v_{\mbox{\boldmath \scriptsize $k$}} $ are subject to the condition,
$
  u_{\mbox{\boldmath \scriptsize $k$}}^2 +
  v_{\mbox{\boldmath \scriptsize $k$}}^2 = 1 $.

In the following analysis, it is convenient to introduce the
two-component operator 
$ \phi_{\mbox{\boldmath \scriptsize $k$}} =
  ( \alpha_{\mbox{\boldmath \scriptsize $k$}},
    \beta^{\dagger}_{-\mbox{\boldmath \scriptsize $k$}}) $ and the 
$ 2 \times 2 $ unit matrix $ \mbox{\boldmath $\tau$}_0 $ and Pauli
matrices $ \mbox{\boldmath $\tau$}_j $ $(j = 1,2,3)$.
The Hamiltonian
$ H =  \tilde{H}_{\rm eh} + \tilde{H}_{\rm pump} $ is expressed
in terms of $ \phi_{\mbox{\boldmath \scriptsize $k$}} $ as follows,
\begin{eqnarray}
  \label{eq:h3}
  H =
    \sum_{\mbox{\boldmath \scriptsize $k$}}
      {\cal E}^{\mu}_{\mbox{\boldmath \scriptsize $k$}}
      \left(
        \phi^{\dagger}_{\mbox{\boldmath \scriptsize $k$}}
        \mbox{\boldmath $\tau$}_{\mu}
        \phi_{\mbox{\boldmath \scriptsize $k$}}
     \right)
  + \frac{1}{2}
    \sum_{\mbox{\boldmath \scriptsize $k$},
          \mbox{\boldmath \scriptsize $p$},
          \mbox{\boldmath \scriptsize $q$}}
      W^{\mu, \nu}_{
         \mbox{\boldmath \scriptsize $k$},
         \mbox{\boldmath \scriptsize $p$}}
       (\mbox{\boldmath $q$})
      \left(
        \phi^{\dagger}_{\mbox{\boldmath \scriptsize $k$}+
                        \mbox{\boldmath \scriptsize $q$}}
        \mbox{\boldmath $\tau$}_{\mu}
        \phi_{\mbox{\boldmath \scriptsize $k$}}
      \right)
      \left(
        \phi^{\dagger}_{\mbox{\boldmath \scriptsize $p$}-
                        \mbox{\boldmath \scriptsize $q$}}
        \mbox{\boldmath $\tau$}_{\nu}
        \phi_{\mbox{\boldmath \scriptsize $p$}}
      \right) ,
\end{eqnarray}
where the summation convention with respect to the indices 
$ \mu, \nu = 0, 1, \cdots, 3 $ is employed.
Introducing the coherence factors, 
$ C^{(\mu)}_{\mbox{\boldmath \scriptsize $k$},
           \mbox{\boldmath \scriptsize $p$}} $
($ \mu = 0, 1, \cdots, 3 $), given by
\begin{eqnarray}
  \label{eq:coh-fact}
  C^{(0)}_{\mbox{\boldmath \scriptsize $k$},
           \mbox{\boldmath \scriptsize $p$}} & = &
  u_{\mbox{\boldmath \scriptsize $k$}}
  u_{\mbox{\boldmath \scriptsize $p$}} +
  v_{\mbox{\boldmath \scriptsize $k$}}
  v_{\mbox{\boldmath \scriptsize $p$}} ,
\nonumber  \\
  C^{(1)}_{\mbox{\boldmath \scriptsize $k$},
           \mbox{\boldmath \scriptsize $p$}} & = &
  u_{\mbox{\boldmath \scriptsize $k$}}
  v_{\mbox{\boldmath \scriptsize $p$}} +
  v_{\mbox{\boldmath \scriptsize $k$}}
  u_{\mbox{\boldmath \scriptsize $p$}} ,
\nonumber   \\
  C^{(2)}_{\mbox{\boldmath \scriptsize $k$},
           \mbox{\boldmath \scriptsize $p$}} & = &
  u_{\mbox{\boldmath \scriptsize $k$}}
  u_{\mbox{\boldmath \scriptsize $p$}} -
  v_{\mbox{\boldmath \scriptsize $k$}}
  v_{\mbox{\boldmath \scriptsize $p$}} , 
\nonumber  \\
  C^{(3)}_{\mbox{\boldmath \scriptsize $k$},
           \mbox{\boldmath \scriptsize $p$}} & = &
  u_{\mbox{\boldmath \scriptsize $k$}}
  v_{\mbox{\boldmath \scriptsize $p$}} -
  v_{\mbox{\boldmath \scriptsize $k$}}
  u_{\mbox{\boldmath \scriptsize $p$}} ,
\end{eqnarray}
and $ {\cal E}^{\mu}_{\mbox{\boldmath \scriptsize $k$}} $ can be
expressed as follows, 
\begin{eqnarray}
  \label{eq:E}
  {\cal E}_{\mbox{\boldmath \scriptsize $k$}}^{0}
  & = &
    \frac{1}{2}
    (\xi_{\mbox{\boldmath \scriptsize $k$}}^{(e)} - 
     \xi_{\mbox{\boldmath \scriptsize $k$}}^{(h)})  ,
\nonumber \\
  {\cal E}_{\mbox{\boldmath \scriptsize $k$}}^{1}
  & = &
    \lambda
    C^{(2)}_{\mbox{\boldmath \scriptsize $k$},
             \mbox{\boldmath \scriptsize $k$}}
   + 
    \frac{1}{2}
    (\xi_{\mbox{\boldmath \scriptsize $k$}}^{(e)} + 
     \xi_{\mbox{\boldmath \scriptsize $k$}}^{(h)} -
     \sum_{\mbox{\boldmath \scriptsize $q$}}
         V_{\mbox{\boldmath \scriptsize $q$}})
    C^{(1)}_{\mbox{\boldmath \scriptsize $k$},
             \mbox{\boldmath \scriptsize $k$}}  ,
\nonumber   \\
  {\cal E}_{\mbox{\boldmath \scriptsize $k$}}^{2} & = & 0  ,
\nonumber   \\
  {\cal E}_{\mbox{\boldmath \scriptsize $k$}}^{3}
  & = &
    \frac{1}{2}
    (\xi_{\mbox{\boldmath \scriptsize $k$}}^{(e)} + 
     \xi_{\mbox{\boldmath \scriptsize $k$}}^{(h)} -
     \sum_{\mbox{\boldmath \scriptsize $q$}}
         V_{\mbox{\boldmath \scriptsize $q$}})
    C^{(2)}_{\mbox{\boldmath \scriptsize $k$},
             \mbox{\boldmath \scriptsize $k$}}
   - 
    \lambda
    C^{(1)}_{\mbox{\boldmath \scriptsize $k$},
             \mbox{\boldmath \scriptsize $k$}}  .
\end{eqnarray}
The quantity 
$ W^{\mu, \nu}_{
     \mbox{\boldmath \scriptsize $k$},
     \mbox{\boldmath \scriptsize $p$}}
   (\mbox{\boldmath $q$})
$
is written as follows,
\begin{eqnarray}
  W^{\mu, \nu}_{
     \mbox{\boldmath \scriptsize $k$},
     \mbox{\boldmath \scriptsize $p$}}
    (\mbox{\boldmath $q$}) = 
  V_{\mbox{\boldmath \scriptsize $q$}}
  \left(
    \begin{array}{cccc}
      C^{(0)}_{\mbox{\boldmath \scriptsize $k$}+
               \mbox{\boldmath \scriptsize $q$},
               \mbox{\boldmath \scriptsize $k$}}
      C^{(0)}_{\mbox{\boldmath \scriptsize $p$}-
               \mbox{\boldmath \scriptsize $q$},
               \mbox{\boldmath \scriptsize $p$}}   &   0   &
    i C^{(0)}_{\mbox{\boldmath \scriptsize $k$}+
               \mbox{\boldmath \scriptsize $q$},
               \mbox{\boldmath \scriptsize $k$}}
      C^{(3)}_{\mbox{\boldmath \scriptsize $p$}-
               \mbox{\boldmath \scriptsize $q$},
               \mbox{\boldmath \scriptsize $p$}}   &  0  \\
      0 & 0 & 0 & 0                                 \\
    i C^{(3)}_{\mbox{\boldmath \scriptsize $k$}+
               \mbox{\boldmath \scriptsize $q$},
               \mbox{\boldmath \scriptsize $k$}}
      C^{(0)}_{\mbox{\boldmath \scriptsize $p$}-
               \mbox{\boldmath \scriptsize $q$},
               \mbox{\boldmath \scriptsize $p$}}   &  0  &
    - C^{(3)}_{\mbox{\boldmath \scriptsize $k$}+
               \mbox{\boldmath \scriptsize $q$},
               \mbox{\boldmath \scriptsize $k$}}
      C^{(3)}_{\mbox{\boldmath \scriptsize $p$}-
               \mbox{\boldmath \scriptsize $q$},
               \mbox{\boldmath \scriptsize $p$}}   &  0  \\
      0 & 0 & 0 & 0                                 
    \end{array}
  \right)_{\mu,\nu}   .
\end{eqnarray}

The Bogoliubov parameters are determined by the variational
method.
We minimize the expectation value of the Hamiltonian,
$ \langle H \rangle_0 $, under the charge neutrality condition,
$ \sum_{\mbox{\boldmath \scriptsize $k$}} \{
  \langle
     c_{\mbox{\boldmath \scriptsize $k$}}^{\dagger}
     c_{\mbox{\boldmath \scriptsize $k$}}
  \rangle_0 +
  \langle
     d_{-\mbox{\boldmath \scriptsize $k$}}^{\dagger}
     d_{-\mbox{\boldmath \scriptsize $k$}}
  \rangle_0 \} = 0
$,
which gives
\begin{eqnarray}
\label{eq:defuv}
  u_{\mbox{\boldmath \scriptsize $k$}}^2 =
  \frac{1}{2} \left(
    1 + \frac{\zeta_{\mbox{\boldmath \scriptsize $k$}}}
             {E_{\mbox{\boldmath \scriptsize $k$}}}
  \right) ,
  v_{\mbox{\boldmath \scriptsize $k$}}^2 =
  \frac{1}{2} \left(
    1 - \frac{\zeta_{\mbox{\boldmath \scriptsize $k$}}}
             {E_{\mbox{\boldmath \scriptsize $k$}}}
  \right) .
\end{eqnarray}
Since 
$ \langle c_{\mbox{\boldmath \scriptsize $k$}}^\dagger
          c_{\mbox{\boldmath \scriptsize $k$}} \rangle_0 = 
  \langle d_{-\mbox{\boldmath \scriptsize $k$}}^\dagger
          d_{-\mbox{\boldmath \scriptsize $k$}} \rangle_0 = 
  v_{\mbox{\boldmath \scriptsize $k$}}^2 $,
the physical meaning of $ v_{\mbox{\boldmath \scriptsize $k$}}^2 $ is
the distribution function of electrons and holes.
In the above equation, $ \zeta_{\mbox{\boldmath \scriptsize $k$}} $,
$ \Delta_{\mbox{\boldmath \scriptsize $k$}} $, and
$ E_{\mbox{\boldmath \scriptsize $k$}} \equiv
  \sqrt{\zeta_{\mbox{\boldmath \scriptsize $k$}}^2 +
        \Delta_{\mbox{\boldmath \scriptsize $k$}}^2}
$
are the renormalized energy of electron-hole pair, the BCS-like energy
gap, and the single particle excitation energy of Bogolon,
respectively. 
The quantities $ \zeta_{\mbox{\boldmath \scriptsize $k$}} $ and
$ \Delta_{\mbox{\boldmath \scriptsize $k$}} $ satisfy the following
self-consistent equations
\begin{mathletters}
\label{eq:gap-eq}
\begin{eqnarray}
  \label{eq:gap-eq1}
  \zeta_{\mbox{\boldmath \scriptsize $k$}}
 & = &
  \left( \frac{k^2}{2 m^{\ast}} - \omega_{\rm L} + E_g \right)
    - 2 \sum_{\mbox{\boldmath \scriptsize $p$}}
      V_{\mbox{\boldmath \scriptsize $k$}-
         \mbox{\boldmath \scriptsize $p$}}
      v_{\mbox{\boldmath \scriptsize $p$}}^2
\nonumber  \\
 & = &
  \left( \frac{k^2}{2 m^{\ast}} - \omega_{\rm L} + E_g \right)
    - \sum_{\mbox{\boldmath \scriptsize $p$}}
      V_{\mbox{\boldmath \scriptsize $k$}-
         \mbox{\boldmath \scriptsize $p$}}
      \left(
         1 - \frac{\zeta_{\mbox{\boldmath \scriptsize $p$}}}
                  {E_{\mbox{\boldmath \scriptsize $p$}}}
      \right) ,  \\
  \label{eq:gap-eq2}
  \Delta_{\mbox{\boldmath \scriptsize $k$}}
 & = &
  2 \lambda
    - 4 \sum_{\mbox{\boldmath \scriptsize $p$}}
      V_{\mbox{\boldmath \scriptsize $k$}-
         \mbox{\boldmath \scriptsize $p$}}
      u_{\mbox{\boldmath \scriptsize $p$}}
      v_{\mbox{\boldmath \scriptsize $p$}}
\nonumber  \\
 & = &
  2 \lambda
    - 2 \sum_{\mbox{\boldmath \scriptsize $p$}}
      V_{\mbox{\boldmath \scriptsize $k$}-
         \mbox{\boldmath \scriptsize $p$}}
      \frac{\Delta_{\mbox{\boldmath \scriptsize $p$}}}
           {E_{\mbox{\boldmath \scriptsize $p$}}} ,
\end{eqnarray}
\end{mathletters}
where $ m^{\ast} = m_e m_h/(m_e + m_h) $ is the reduced mass of
electron-hole pairs.
The second term of the right-hand side of Eq.~(\ref{eq:gap-eq1})
expresses the band renormalization effect arising from the electron
(hole) exchange interaction.
The Eq.~(\ref{eq:gap-eq2}) is the BCS-like gap equation where the
BCS-like energy gap formation induced by the pump-light is taken into 
account. 

In order to obtain the physical insight into the BCS-like gap equation,
we express the BCS-like gap equation, Eq.~(\ref{eq:gap-eq}), in terms
of the electron-hole pair wave function 
with zero center-of-mass momentum \cite{NSR},
$ \psi_{\mbox{\boldmath \scriptsize $k$}} = 
  \langle c_{\mbox{\boldmath \scriptsize $k$}}
          d_{-\mbox{\boldmath \scriptsize $k$}}
  \rangle =
  \Delta_{\mbox{\boldmath \scriptsize $k$}}/
  (2 E_{\mbox{\boldmath \scriptsize $k$}}) .
$
Making use of Eq.~(\ref{eq:defuv}), the BCS-like gap equation can be
rewritten as
\begin{eqnarray}
  \label{eq:Wannier}
  \left\{
    \frac{k^2}{2 m^{\ast}} - (\omega_{\rm L} - E_g) -
    2 \sum_{\mbox{\boldmath \scriptsize $p$}}
      V_{\mbox{\boldmath \scriptsize $k$}-
         \mbox{\boldmath \scriptsize $p$}}
      v_{\mbox{\boldmath \scriptsize $p$}}^2
  \right\} \psi_{\mbox{\boldmath \scriptsize $k$}}
  - (1 - 2 v_{\mbox{\boldmath \scriptsize $k$}}^2)
  \sum_{\mbox{\boldmath \scriptsize $p$}}
      V_{\mbox{\boldmath \scriptsize $k$}-
         \mbox{\boldmath \scriptsize $p$}}
      \psi_{\mbox{\boldmath \scriptsize $p$}}
  = (1 - 2 v_{\mbox{\boldmath \scriptsize $k$}}^2) \lambda .
\end{eqnarray}
This equation is reduced to the Wannier equation in the limit of low
electron-hole densities and $ \lambda \rightarrow 0 $, because
$ v_{\mbox{\boldmath \scriptsize $k$}}^2 $ is the pair distribution
function .
Therefore the BCS-like pair theory is able to properly describe the
relative wave-function of excitons in the low density case as well as 
the BCS-like pair states in the high-density case.

\subsection{The Bethe-Salpeter equation for electron-hole pair Green
  function} 
As described above, the optical spectrum is obtained by the evaluation
of the electron-hole retarded Green function $ G_{\rm R}(t) $, which
can be rewritten in terms of Bogoliubov parameters as,
\begin{eqnarray}
  \label{eq:2-5}
  G_{\rm R}(t) =
  \sum_{\mbox{\boldmath \scriptsize $k$},
        \mbox{\boldmath \scriptsize $p$}}
    g_{\mbox{\boldmath \scriptsize $k$}}^{\ast}
    g_{\mbox{\boldmath \scriptsize $p$}}
    \mbox{\boldmath $K$}_{\mbox{\boldmath \scriptsize $k$}}
    \mbox{\boldmath ${\cal G}$}_{\mbox{\boldmath \scriptsize $k$},
                                 \mbox{\boldmath \scriptsize $p$}}(t)
    \mbox{\boldmath $K$}_{\mbox{\boldmath \scriptsize $p$}} ,
\end{eqnarray}
where
$ \mbox{\boldmath $K$}_{\mbox{\boldmath \scriptsize $k$}} \equiv (
   u_{\mbox{\boldmath \scriptsize $k$}}^2 -
   v_{\mbox{\boldmath \scriptsize $k$}}^2 ,
   i) $
is the two-component vector.
$ \mbox{\boldmath ${\cal G}$}_{
    \mbox{\boldmath \scriptsize $k$},
    \mbox{\boldmath \scriptsize $p$}} (t) $
is the $ 2 \times 2 $ matrix retarded Green function whose
$(\mu, \nu)$ component is given by 
\begin{eqnarray}
  \label{eq:3-1b}
 \biggl(
  \mbox{\boldmath ${\cal G}$}_{
    \mbox{\boldmath \scriptsize $k$},
    \mbox{\boldmath \scriptsize $p$}} (t)
  \biggr)_{\mu, \nu} =
  - i \Theta(t) \langle [
  ( \phi_{\mbox{\boldmath \scriptsize $k$}}^{\dagger}(t)
    \mbox{\boldmath $\tau$}_{\mu}
    \phi_{\mbox{\boldmath \scriptsize $k$}}(t) ),
  ( \phi_{\mbox{\boldmath \scriptsize $p$}}^{\dagger}(0)
    \mbox{\boldmath $\tau$}_{\nu}
    \phi_{\mbox{\boldmath \scriptsize $p$}}(0) )
  ] \rangle ,
\end{eqnarray}
where $ \mu, \nu = 1, 2 $.
The function
$ \mbox{\boldmath ${\cal G}$}_{
    \mbox{\boldmath \scriptsize $k$}, 
    \mbox{\boldmath \scriptsize $p$}} (\omega) $
satisfies the following BS equation,
\begin{eqnarray}
  \label{eq:bs-eq}
  (\omega \mbox{\boldmath $\tau$}_0
  + E_{\mbox{\boldmath \scriptsize $k$}} \mbox{\boldmath $\tau$}_2) 
  \mbox{\boldmath ${\cal G}$}_{
    \mbox{\boldmath \scriptsize $k$},
    \mbox{\boldmath \scriptsize $p$}} (\omega)
  - \sum_{\mbox{\boldmath \scriptsize $k$}'}
  V_{\mbox{\boldmath \scriptsize $k$}
    -\mbox{\boldmath \scriptsize $k$}'}
  \left(
    C^{(0)2}_{
      \mbox{\boldmath \scriptsize $k$},
      \mbox{\boldmath \scriptsize $k$}'}
    \mbox{\boldmath $\tau$}_2
  + i
    C^{(3)2}_{
      \mbox{\boldmath \scriptsize $k$},
      \mbox{\boldmath \scriptsize $k$}'}
    \mbox{\boldmath $\tau$}_1
  \right)
  \mbox{\boldmath ${\cal G}$}_{
    \mbox{\boldmath \scriptsize $k$}',
    \mbox{\boldmath \scriptsize $p$}} (\omega)
  = \mbox{\boldmath $\tau$}_2
  \delta_{\mbox{\boldmath \scriptsize $k$},
          \mbox{\boldmath \scriptsize $p$}} .
\end{eqnarray}
This BS equation is obtained by linearizing the equation-of-motion for
$ \mbox{\boldmath ${\cal G}$}_{
    \mbox{\boldmath \scriptsize $k$},
    \mbox{\boldmath \scriptsize $p$}}(t) $.
The second term on the left-hand side of Eq.~(\ref{eq:bs-eq}) describes
the effect of multiple Coulomb interaction between Bogolons, which is
attractive in the low electron-hole density limit.

The vertex part proportional to 
$ C^{(0)2}_{\mbox{\boldmath \scriptsize $k$},
            \mbox{\boldmath \scriptsize $p$}} $ 
reflects the state-filling effect and the part proportional to
$ C^{(3)2}_{\mbox{\boldmath \scriptsize $k$},
            \mbox{\boldmath \scriptsize $p$}} $ 
arises because of the electron-hole pair correlation.
It should be emphasized that this term describes the deviation from
the BCS-like mean-field theory, and give rise to the sharp excitonic
structures in the pump-probe spectrum. 

\subsection{Screening effect}
We consider the screening effect within the quasi-static RPA following
the work by Haug and Schmitt-Rink \cite{Haug}.
The BCS-like gap equation, Eq.~(\ref{eq:gap-eq}), is rewritten as
follows, 
\begin{mathletters}
\label{eq:s-gap-eq}
\begin{equation}
  \label{eq:s-gap-eq1}
  \zeta_{\mbox{\boldmath \scriptsize $k$}} =
  \left(
    \frac{k^2}{2 m^{\ast}} - \omega_{\rm L} + E_g
  \right)
   - 2 \sum_{\mbox{\boldmath \scriptsize $p$}}
      V^{\rm s}_{\mbox{\boldmath \scriptsize $k$}-
         \mbox{\boldmath \scriptsize $p$}}
      v_{\mbox{\boldmath \scriptsize $p$}}^2
    + \sum_{\mbox{\boldmath \scriptsize $q$}}
      \left\{
         V^{\rm s}_{\mbox{\boldmath \scriptsize $q$}}
       - V_{\mbox{\boldmath \scriptsize $q$}}
      \right\} ,
\end{equation}
\begin{equation}
  \label{eq:s-gap-eq2}
  \Delta_{\mbox{\boldmath \scriptsize $k$}} =
  2 \lambda
    - 4 \sum_{\mbox{\boldmath \scriptsize $p$}}
      V^{\rm s}_{\mbox{\boldmath \scriptsize $k$}-
         \mbox{\boldmath \scriptsize $p$}}
      u_{\mbox{\boldmath \scriptsize $p$}}
      v_{\mbox{\boldmath \scriptsize $p$}} ,
\end{equation}
\end{mathletters}
where $ V^{\rm s}_{\mbox{\boldmath \scriptsize $q$}} $ is the
quasi-statically screened Coulomb potential.
Here the second term on the right-hand side of Eq.~(\ref{eq:s-gap-eq1})
represent the screened exchange effect and the third term being the
Coulomb-hole effect.
The BS equation, Eq.~(\ref{eq:bs-eq}), is rewritten as
\begin{eqnarray}
  \label{eq:bs-eq2}
  (\omega \mbox{\boldmath $\tau$}_0
  + E_{\mbox{\boldmath \scriptsize $k$}} \mbox{\boldmath $\tau$}_2) 
  \mbox{\boldmath ${\cal G}$}_{
    \mbox{\boldmath \scriptsize $k$},
    \mbox{\boldmath \scriptsize $p$}} (\omega)
  - \sum_{\mbox{\boldmath \scriptsize $k$}'}
  V^{\rm s}_{\mbox{\boldmath \scriptsize $k$}
    -\mbox{\boldmath \scriptsize $k$}'}
  \left(
    C^{(0)2}_{
      \mbox{\boldmath \scriptsize $k$},
      \mbox{\boldmath \scriptsize $k$}'}
    \mbox{\boldmath $\tau$}_2
  + i
    C^{(3)2}_{
      \mbox{\boldmath \scriptsize $k$},
      \mbox{\boldmath \scriptsize $k$}'}
    \mbox{\boldmath $\tau$}_1
  \right)
  \mbox{\boldmath ${\cal G}$}_{
    \mbox{\boldmath \scriptsize $k$}',
    \mbox{\boldmath \scriptsize $p$}} (\omega)
  = \mbox{\boldmath $\tau$}_2
  \delta_{\mbox{\boldmath \scriptsize $k$},
          \mbox{\boldmath \scriptsize $p$}} .
\end{eqnarray}

We employ the tractable expression for the dielectric function given
by the single-plasmon-pole approximation, which is known to produce
the relatively good self-energy corrections \cite{Iida,LSH}
The explicit expression for the screened potential is given by,
\begin{eqnarray}
  \label{eq:spp}
  V^{\rm s}_{\mbox{\boldmath \scriptsize $q$}} =
  \frac{4 \pi e^2}{\epsilon_{\mbox{\boldmath \scriptsize $q$}} q^2} ,
  \epsilon_{\mbox{\boldmath \scriptsize $q$}}^{-1} =
  \epsilon_0^{-1}
    \left(
      1 - \frac{\omega_{\rm pl}^2}
               {\omega_{\mbox{\boldmath \scriptsize $q$}}^2}
    \right) ,
\end{eqnarray}
where $ \omega_{\rm pl} = 4 \pi n e^2/(\epsilon_0 m^{\ast}) $ is the
plasma frequency, and $ n $ is the electron-hole pair density.
The dispersion of the effective plasmon mode is given by \cite{Zimmermann}
\begin{eqnarray}
  \label{eq:spp2}
  \omega_{\mbox{\boldmath \scriptsize $q$}}^2 =
  \omega_{\rm pl}^2
  \left( 1 + \frac{q^2}{k_{\rm TF}^2} \right) +
  \left( \frac{q^2}{4 m^{\ast}} \right)^2 +
  G_{\rm eff}^2 ,
\end{eqnarray}
where 
$ k_{\rm TF} = 
  \{ 16 m^{\ast} e^2/(\pi \epsilon_0) \}(6 \pi^2 n)^{-1/3} $
is the Thomas-Fermi wave number, and the effective gap $ G_{\rm eff} $
is set to the minimum energy of the Bogolon pair excitation.

\section{Numerical Analysis}
\label{sec:3}
\subsection{%
  Solution of the BCS-like gap equation and the
  resonatorless optical bistability}
\label{sec:3-1}

In the following analysis, we use the units where the exciton binding
energy and the exciton Bohr radius being unity. 
For simplicity, we assume that the electron and hole masses are
identical, i.e., $ m_e = m_h $.

As a first step in our numerical calculations, we iteratively solve
the BCS-like gap equation, Eq.~(\ref{eq:s-gap-eq}).
Here we note that the quasi-stationary state of this optically driven
electron-hole system is controlled by changing $ \omega_{\rm L} $ and
$ \lambda $. 

We first show the calculated mean interparticle distance, 
$ r_{\rm s} = (4 \pi n/3)^{-1/3} $, as a function of pump-frequency in
Fig.~\ref{fig:1}, where $ n $ is the electron-hole density.
We find the resonatorless optical bistability in which the two stable
states exist for the same pump-frequency, as shown in 
Ref.~\onlinecite{Iida}.
With increasing the interaction energy $ \lambda $ between carriers and 
pump-light, the bistability region of $ \omega_{\rm L} $ becomes
narrow and, the bistable behavior disappears for 
$ \lambda \gtrsim 0.3 $.

For convenience in the following discussion, we show in 
Fig.~\ref{fig:Geff2} the pump-frequency $ \omega_{\rm L} $ dependence
of the effective-gap $ G_{\rm eff} $, which is defined by the minimum
value of the single-particle excitation energy  
$ E_{\mbox{\boldmath \scriptsize $k$}} $.
In the high electron-hole density generated by the resonant
excitations ($ \omega_{\rm L} - E_g > 0 $), the weak electron-hole
pair correlation predominates, in which the BCS-like energy gap is
formed at the quasi-Fermi level.
With decreasing the pump-frequency, the BCS-like gap 
$ \Delta_{\mbox{\boldmath \scriptsize $k$}} $ continues to exist even
for the off-resonance cases especially in the higher density side of
the bistable region.
This is because the system undergoes the resonance condition because
of the band-gap renormalization effect.
We should remind that the BCS-like gap for degenerate electron-hole
systems is much smaller than the exciton binding energy.
The special attention should be paid to the fact that the magnitude of
the BCS-like gap becomes very large with increasing $ \lambda $, which
is shown in Fig.~\ref{fig:Geff2}.
This behavior arises because the electron-hole pair correlation is
strongly enhanced by the coherent pump-light, which was pointed out in
Ref.~\onlinecite{Comte2} as the ``light enhanced electron-hole
correlation''. 
On the contrary, in the low electron-hole density generated by the
large off-resonant excitations ($ \omega_{\rm L} - E_g \ll 0 $), the
contribution of the Coulomb interaction to $ G_{\rm eff} $ is
disregarded, and $ G_{\rm eff} $ is approximated to 
$ E_g - \omega_{\rm L} $.

\subsection{%
  Pump-probe spectra under weak photoexcitations}
\label{sec:3-2}
Now, let us consider the pump-probe spectrum $ A(\omega) $ given by
Eq.~(\ref{eq:pp}).
The positive and negative values of $ A(\omega) $ imply the absorption
and the stimulated emission of the probe light, respectively. 
When we numerically solve the BS equation, Eq.~(\ref{eq:bs-eq2}),
the usual singularity-removal method \cite{HK} is inappropriate
because the singularity of the retarded Green function considerably
depends on the electron-hole density.
This fact forces us to deal with a giant matrix 
(about $ 3000 \times 3000 $ in dimension) to obtain good overall
accuracy. 
In the present analysis we numerically evaluate the eigenvalues and
eigenvectors of the stability matrix \cite{Ring} of 
the BS kernel in Eq.~(\ref{eq:bs-eq2}).

In Figs.~\ref{fig:2}-\ref{fig:4}, we show the calculated spectra for 
several values of $ \omega_{\rm L} $ and $ \lambda = 0.001 $, in which
the bistability region exists. 
Figure~\ref{fig:2} is the spectrum for a higher density state 
($ \omega_{\rm L} - E_g = -0.5 $, $ r_{\rm s} =0.87 $) outside the
bistability region, and Fig.~\ref{fig:3} is that for a lower density
state ($ \omega_{\rm L} - E_g = -4.0 $, $ r_{\rm s} = 140 $)
outside the bistability region.
The spectra inside the bistability region
($ \omega_{\rm L} - E_g = -2.5 $) in higher ($ r_{\rm s} = 1.0 $) and
lower ($ r_{\rm s} = 86 $) density states are shown in
Fig.~\ref{fig:4}. 

In the high-density state (Fig.~\ref{fig:2}) outside the bistability
region, the strong stimulated emission, and the large red-shift of the
band-edge due to the band renormalization effect is found.
The transparent region arising from the BCS-like energy gap formation
\cite{Iida} is almost vanishing in this choice of parameters, because
the BCS-like gap is extremely small as shown in Fig.~\ref{fig:Geff2}. 
However, this does not imply that the BCS-like gap is not formed;
actually, around $ \omega \simeq \omega_{\rm L} $, we find a pair of
sharp peaked structures in the pump-probe spectrum, which originates
from the singular behavior of the density-of-states accompanied by
the BCS-like energy gap formation due to the many-body effect. 
It should be noted that this anomaly is weak in the BCS-like
mean-field calculation as shown by the dashed line in
Fig.~\ref{fig:2}, and the BS equation, Eq.~(\ref{eq:bs-eq2}), properly
describes the strong electron-hole pair correlation.

In contrast to Fig.~\ref{fig:2}, the pump-probe spectrum in the
low-density state outside the bistability region exhibits the
qualitatively different profile. 
We find, in Fig.~\ref{fig:3}, a series of sharp absorption lines
corresponding to the 1S, 2S $\cdots$ exciton states, and the
calculated spectrum is reduced to the one given by the Elliott formula
\cite{Haug} in the low density limit.
However, it should be noted that the present analysis is based on the
electron-hole many-body theory including the strong electron-hole pair
correlation, and the large deviations from the square-root dependence
of  $ A(\omega) $ is observed close to the band-edge.
In addition, it should be stressed that the BCS-like decoupling
approximation for electron-hole pair Green function cannot reproduce
the excitonic structure, even though the BCS-like gap equation
properly describes the internal motion of excitons both in the high
and low density limits.

Inside the bistability region, the high- and the low-density behavior
appears for the same pump frequency $ \omega_{\rm L} $, as shown in
Fig.~\ref{fig:4}.
We should recall that this bistability arises from the instability due
to the positive feedback between the optical pumping and the band-gap
renormalization. 

\subsection{%
  The dependence of the pump-probe spectra on the pump-light
  intensity} 
\label{sec:3-3}
In the following, let us discuss the dependence of the pump-probe
spectrum on the interaction energy between carriers and pump-light
$ \lambda $.
We show in Fig.~\ref{fig:L1} the calculated spectra for three states:
$ \lambda = 0.1, 0.2 $ and $ 0.3 $.
Here the pump-light frequency is chosen as 
$ \omega_{\rm L} - E_g = -0.5 $.

As in the case of Fig.~\ref{fig:2}, we find the stimulated emission
in $ \omega < \omega_{\rm L} $, and the absorption components in 
$ \omega > \omega_{\rm L} $.
As shown in Fig.~\ref{fig:1}, the mean interparticle distance is 
$ r_{\rm s} \approx 0.87 $, irrespective of the magnitude of 
$ \lambda $, because the electron-hole recombination rate is assumed
to be much smaller than the Rabi frequency.
Between the absorption and the emission parts, we find the transparent
region originating from the BCS-like gap formation, which is extremely
enhanced by the pump-light excitation.
Conversely, the Rabi splitting of the dressed semiconductor states by
the strong pump light is combined with the electron-hole Coulomb
interaction to form extremely large BCS-like gap.
In the large off-resonance case (and the low-density state in the
bistability region), the energy gap becomes insensitive to 
$ \lambda $, as shown in Fig.~\ref{fig:Geff2}.

With increasing $ \lambda $, the broad absorption (emission) component 
splits into a sharp absorption (emission) line and the broad
absorption (emission) band.
In the following discussion, we call these sharp stimulated emission
and the absorption lines as 
$ A_{\rm e} $-line and $ A_{\rm a} $-line, respectively.
The $ A_{\rm e} $- and $ A_{\rm a} $-lines appear inside the
transparent region given by the BCS-like mean field calculation which
is shown by the dashed curves in Figs.~\ref{fig:2}-\ref{fig:4} and
Figs.~\ref{fig:8}-\ref{fig:5}.
The $ A_{\rm e}$- and $ A_{\rm a} $-lines are manifestation of the
optical transition accompanied by the excitation of the collective
phase fluctuation mode in the electron-hole BCS state, which
corresponds to the Anderson mode in superconductivity \cite{Anderson}. 
The attention should be paid to the fact that this mode is induced by
the strong pump-light, which is reduced to the band edge singularity
for the weak pump-light as discussed above.
That is to say, these lines do not simply arise from the conventional
excitonic processes, because the exciton picture breaks down in this
density regime and their spectral positions are not located in the
semiconductor band gap but in the BCS-like gap. 
Since the collective phase fluctuation mode is stabilized under the
intense pump-light, the splitting between the $ A_{\rm e(a)} $-line
and the broad stimulated emission (absorption) component becomes
larger with increasing $ \lambda $. 

\subsection{%
  Pump-probe spectra under strong photoexcitations}
\label{sec:3-4}

In the following, we discuss the dependence of the pump-probe spectra
on the pump-light frequency $ \omega_{\rm L} $ in the high-density
states under the strong photoexcitations. 
Figures \ref{fig:8}-\ref{fig:5} depict the spectra for 
$ \lambda = 0.3 $, where the interaction energy between the pump-light 
and carriers is sufficiently strong as to suppress the resonatorless
optical bistability \cite{Iida}.

We show in Fig.~\ref{fig:8} the calculated spectrum for 
$ \omega_{\rm L} - E_g = -1.8 $, in which the mean interparticle
distance is $ r_{\rm s} = 1.0 $. 
In spite of the relatively large off-resonance, we find the 
$ A_{\rm e} $-line ($ A_{\rm a} $-line) in the higher (lower)
frequency side of the broad emission (absorption) component as shown
in Fig.~\ref{fig:L1}; this is because the large band renormalization
preserves the system in the resonance condition.

We next show the spectra for 
$ \omega_{\rm L} - E_g = -2.75 $ (Fig.~\ref{fig:7}),
$ \omega_{\rm L} - E_g = -2.9 $ (Fig.~\ref{fig:6}) and
$ \omega_{\rm L} - E_g = -3.2 $ (Fig.~\ref{fig:5}), where the mean
interparticle distances are $ r_{\rm s} = 1.4 $,
$ r_{\rm s} = 1.8 $ and $ r_{\rm s} = 2.3 $, respectively. 
Making a comparison between Figs.~\ref{fig:8}-\ref{fig:6}, we find
that the broad emission (absorption) component splits into a series of 
sharp emission (absorption) lines and a broad emission (absorption)
band with decreasing the pump-light frequency $ \omega_{\rm L} $.
The spectral positions of the new series of sharp emission lines and
the $ A_{\rm e} $-line slightly shift towards longer wavelength with
decreasing the pump-light frequency, because the effective gap  
$ G_{\rm eff} $ increases as shown in Fig.~\ref{fig:Geff2}.
We find, in Figs.~\ref{fig:6}-\ref{fig:5} that the intensities of this
series of sharp emission lines and the broad emission band become weak 
as the pump-light frequency decreases.
It is pointed out that neither this series nor the 
$ A_{\rm e/a} $-line is obtained by the BCS-like pairing theory even
with the RPA correction.

We should be noted that those series of sharp emission/absorption 
components reflect the crossover behavior from the optically induced
collective fluctuation to the exciton real-space pairing with multiple
bound states (Wannier series). 

In the large off-resonant excitation (Figs.~\ref{fig:6}-\ref{fig:5}),
a series of sharp absorption components including the 
$ A_{\rm a}$-line can be regarded as the exciton lines, where the
excitonic real-space pairing in the many exciton system play the major 
role.
Moreover, the spectral positions for emission components are well
described by the nonlinear optical processes in many exciton systems
(the exciton transition by the combination of the two pump-photon
absorption and one probe-photon emission)
rather than the pair recombination processes accompanied by the
excitation of the collective phase fluctuation mode in the
electron-hole BCS state. 
In fact, we find in Figs.~\ref{fig:6}-\ref{fig:5} that the
magnitude of the effective gap $ G_{\rm eff} $ is approximately equal
to $ E_g - \omega_{\rm L} $, and the frequency ($\omega$) of the 
$ A_{\rm e} $-line and the series of sharp emission lines are given by 
$ 2 \omega_{\rm L} - \omega = E_g - E_{n,x} $, where $ E_{n,x} $ with 
$ n = 1, 2, \cdots $ is the binding energy of the $n$S exciton and 
$ n=1 $ ($ n=2, 3, \cdots $) corresponds to $ A_{\rm e} $-line (the
series of sharp emission lines).

On the contrary, Fig.~\ref{fig:7} can be regarded as the intermediate
state between the light-enhanced electron-hole BCS state and the many
exciton system.
This is because the broad band and the sharp lines coexist in the
stimulated emission spectrum, and the $ A_{\rm e} $-line has the
intermediate character between the collective phase fluctuation and
the above-mentioned nonlinear optical process for the 1S exciton.

We next compare the calculated spectra given by the present theory and
by the BCS-like mean-field theory.
Figures~\ref{fig:7}-\ref{fig:6} indicate that the spectral positions
of the series of sharp emission and absorption lines appear
inside the transparent region given by the BCS-like mean-field
calculation.
It should be noted here that the origin of those series of sharp
emission and absorption lines including $ A_{\rm e} $- and 
$ A_{\rm a} $-lines are different from that of pair of peaked
structures in the BCS-like mean-field calculation, which
originate from the singularity of the density-of-states for the
Bogolon excitation. 
In addition, we find, in Fig.~\ref{fig:7}, that the
intensity of the broad stimulated emission component given by the
present theory is much weaker than that of the corresponding spectral 
component predicted by the BCS-like mean-field calculation, 
and the intensity of the broad emission band is almost vanishing in
Figs.~\ref{fig:6}-\ref{fig:5}.

\section{Conclusion}
\label{sec:4}
\label{sec:conclusion}

We present a many-body theory for pump-probe spectra in highly excited
semiconductors which is applicable throughout the whole electron-hole
density states; the high-density electron-hole BCS state and the low
density excitonic BEC state are contained as the limits. 
The present theory is based on the BCS-like pairing theory combined
with the BS equation for electron-hole pair propagator, which first
enables us to consider the state-filling effect, the band-gap
renormalization and the strong/weak pair correlation in a unified
manner.

We calculate the pump-probe spectra under the weak photoexcitations.
In the high density state generated by a small off-resonant pumping,
we first show that a pair of sharp peaked structure near the
pump-light frequency is considerably enhanced by the strong pair
correlation. 
On the other hand, in the low density state given by a large
off-resonant pumping, we find the excitonic absorption structures near 
the band-edge, which is missing in the BCS-like mean-field theory
even with the RPA correction \cite{Comte2,Iida}.

We next analyze the dependence of the spectra on the pump-light
intensities.
It is shown that the electron-hole BCS state is stabilized and the
BCS-like gap is distinctly enhanced by the intense pump-light.
This result strongly suggests that the macroscopic quantum phenomena,
such as excitonic superfluidity, can be observed under the strong
photoexcitation. 

With increasing the pump-light intensity, we find that those new sharp
emission and absorption lines ($A_{\rm e}$- and $A_{\rm a}$-lines)
split from the broad emission and absorption components, respectively.
These new lines originate from the optical transition accompanied by
the collective phase fluctuation mode in the
electron-hole BCS state, which corresponds to the Anderson mode in
superconductivity \cite{Anderson}. 

Finally, we analyze the dependence of the calculated spectra on the
pump-light frequency under the strong photoexcitation, and show that
the collective phase fluctuation mode continuously changes to the
exciton mode as the pump-light frequency decreases.

\section*{Acknowledgment}
\label{sec:acknowledgment}
This work is partially supported by a Grant-in-Aid for Scientific
Research on priority areas, ``Photo-induced Phase Transition and Their
Dynamics'' from the Ministry of Education, Science, Culture and Sports
of Japan. 



\begin{figure}
\caption{%
  The mean interparticle distance $ r_{\rm s} = (4 \pi n/3)^{-1/3} $
  as a function of pump-light frequency, $ \omega_{\rm L} - E_g $,
  where $ n $ is the electron-hole pair density. 
  The interaction energies between particles and pump light
  $ \lambda $ are chosen (top to bottom): 0.001, 0.1, 0.3, 1.0.
  The open and closed circles represent the point at which the
  pump-probe spectra are calculated for $ \lambda = 0.001 $ and
  $ \lambda = 0.3 $, respectively.
}
\label{fig:1}
\end{figure}

\begin{figure}
\caption{%
  The effective gap $ G_{\rm eff} $ as a function of the pump
  frequency $ \omega_{\rm L} $.
  The interaction energies between particles and pump light
  $ \lambda $ are chosen (bottom to top): 0.001, 0.1, 0.3, 1.0.
}
\label{fig:Geff2}
\end{figure}

\begin{figure}
\caption{%
  The pump-probe spectrum in the high-density state outside the
  bistability region ($ \omega - E_g = -0.5 $, $ r_{\rm s} = 0.87 $). 
  Here the interaction energy between particles and pump light is
  chosen as $ \lambda = 0.001 $.
  (a) is given by the present theory, and (b) is given by the BCS-like
  mean field theory.
}
\label{fig:2}
\end{figure}

\begin{figure}
\caption{%
  The pump-probe spectrum in the low-density state outside the
  bistability region ($ \omega - E_g = -4.0 $, $ r_{\rm s} = 140 $).
  The interaction energy between particles and pump light is
  $ \lambda = 0.001 $.
  (a) is given by the present theory, and (b) is given by the BCS-like
  mean field theory.
}
\label{fig:3}
\end{figure}

\begin{figure}
\caption{%
  The pump-probe spectrum for the states inside the bistability region
  ($ \omega - E_g = -2.0 $).
  Here the interaction energy between particles and pump light is
  chosen as $ \lambda = 0.001 $.
  (a) and (b) are the spectra for low-density state ($ r_{\rm s} = 86 $)
  calculated by the present theory and BCS-like mean field theory,
  respectively.
  (c) and (d) are the results for high-density state 
  ($ r_{\rm s} = 1.0 $) given by the present theory and BCS-like
  mean field theory, respectively.
}
\label{fig:4}
\end{figure}

\begin{figure}
\caption{%
  The pump-probe spectrum for $ \lambda = 0.1, 0.2, 0.3 $, in which
  the pump-light frequency is chosen as 
  $ \omega_{\rm L} - E_g = -0.5 $.
  In this choice of $ \omega_{\rm L} - E_g $, the dimensionless mean
  interparticle distance ($ r_{\rm s} = 0.87 $) is almost independent
  on $ \lambda $ as shown in Fig.~\ref{fig:1}.
}
\label{fig:L1}
\end{figure}

\begin{figure}
\caption{%
  The pump-probe spectrum for $ \omega - E_g = -1.8 $, in which the 
  dimensionless mean interparticle distance is $ r_{\rm s} = 1.0 $.
  The interaction energy between particles and the pump-light is
  $ \lambda = 0.3 $, in which resonatorless optical bistability
  disappears.
  (a) is given by the present theory, and (b) is given by the BCS-like
  mean field theory.
}
\label{fig:8}
\end{figure}

\begin{figure}
\caption{%
  The pump-probe spectrum for $ \omega - E_g = -2.75 $, in which the 
  dimensionless mean interparticle distance is $ r_{\rm s} = 1.4 $.
  The interaction energy between particles and the pump-light is
  $ \lambda = 0.3 $, in which resonatorless optical bistability
  disappears.
  (a) is given by the present theory, and (b) is given by the BCS-like
  mean field theory.
}
\label{fig:7}
\end{figure}

\begin{figure}
\caption{%
  The pump-probe spectrum for $ \omega - E_g = -2.9 $, in which the 
  dimensionless mean interparticle distance is $ r_{\rm s} = 1.8 $.
  The interaction energy between particles and the pump-light is
  $ \lambda = 0.3 $, in which resonatorless optical bistability
  disappears.
  (a) is given by the present theory, and (b) is given by the BCS-like
  mean field theory.
}
\label{fig:6}
\end{figure}

\begin{figure}
\caption{%
  The pump-probe spectrum for $ \omega - E_g = -3.2 $, in which the 
  dimensionless mean interparticle distance is $ r_{\rm s} = 2.3 $. 
  The interaction energy between particles and the pump-light is
  $ \lambda = 0.3 $, in which resonatorless optical bistability
  disappears.
  (a) is given by the present theory, and (b) is given by the BCS-like
  mean field theory.
}
\label{fig:5}
\end{figure}

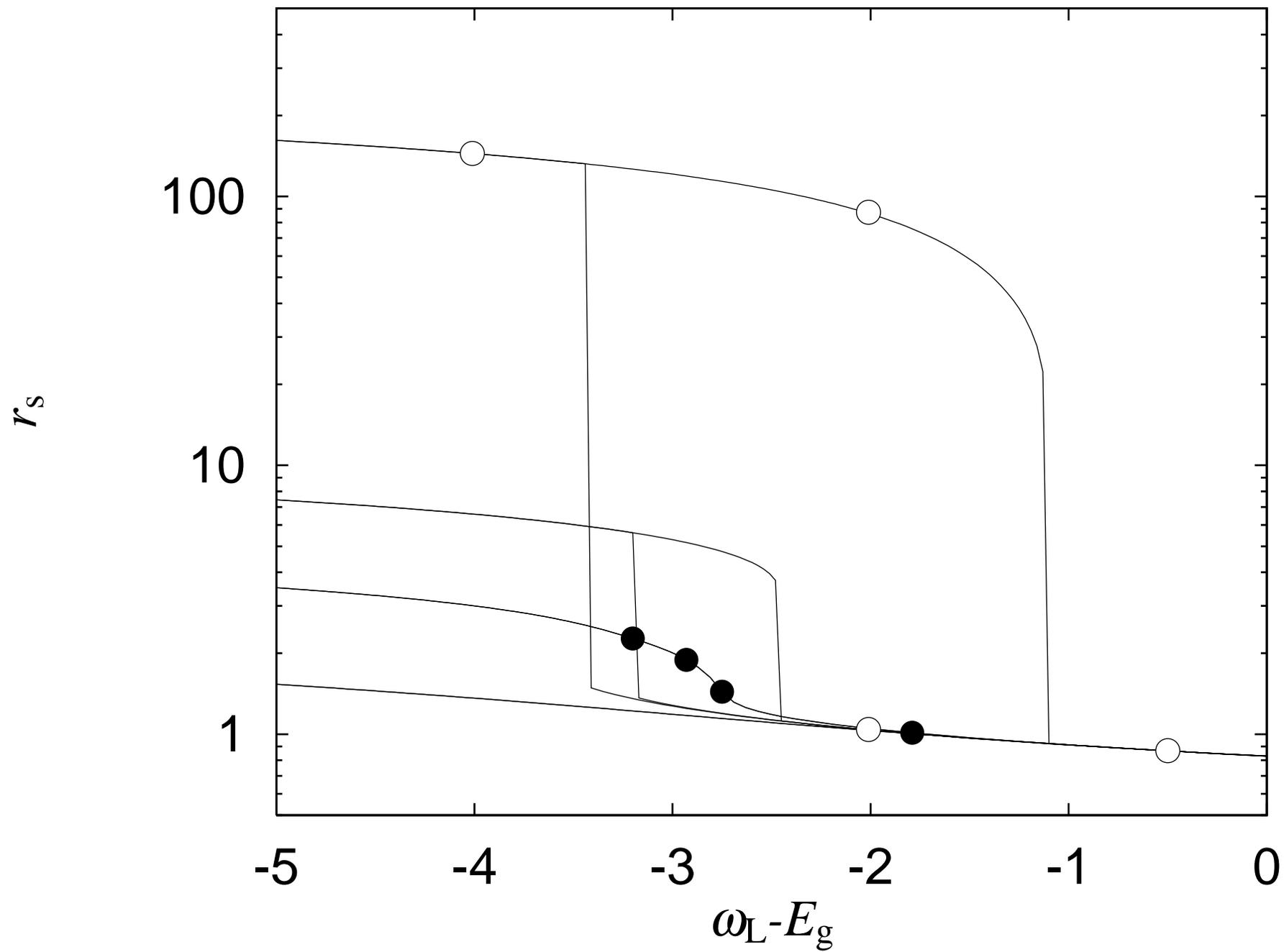
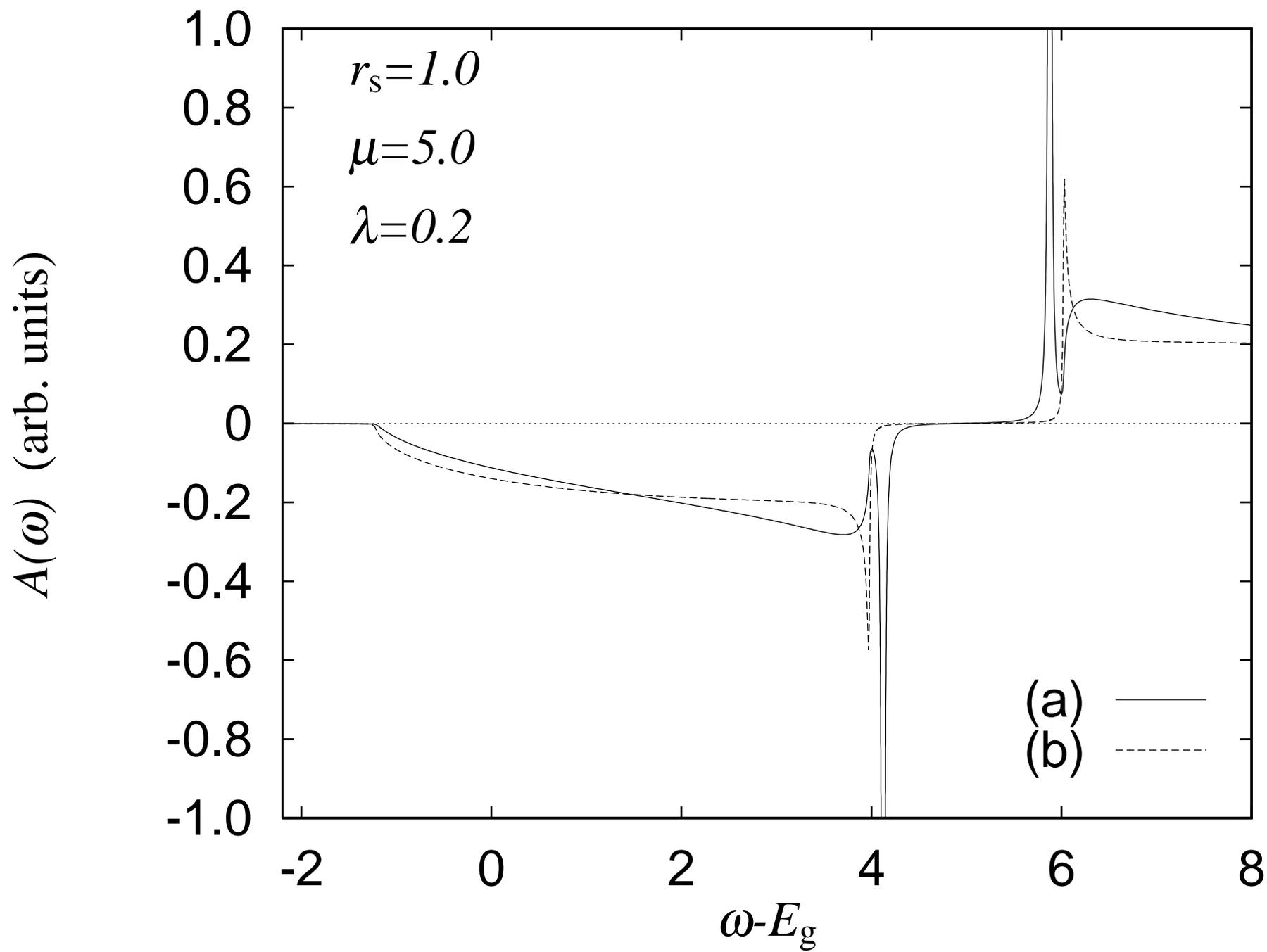
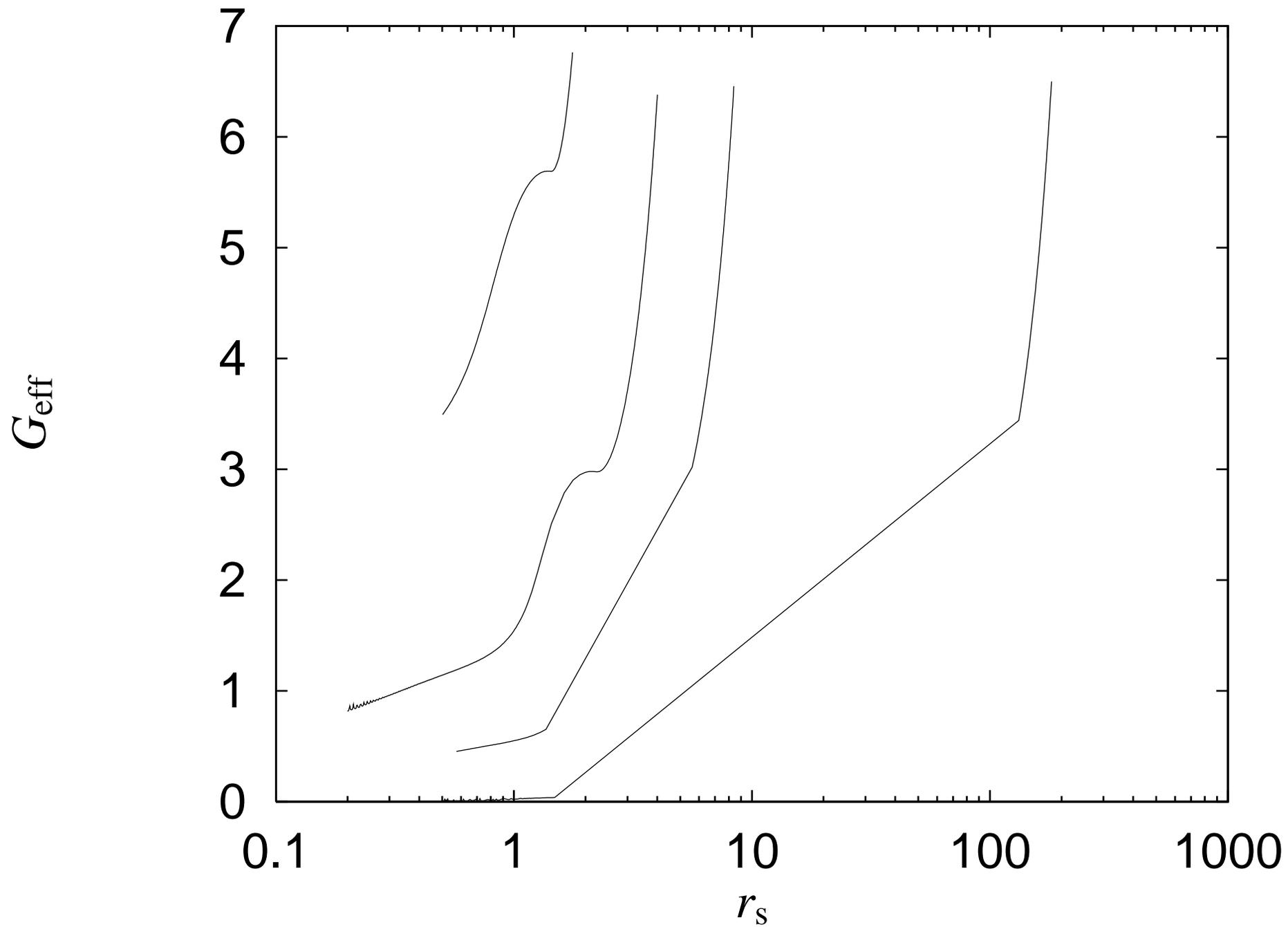
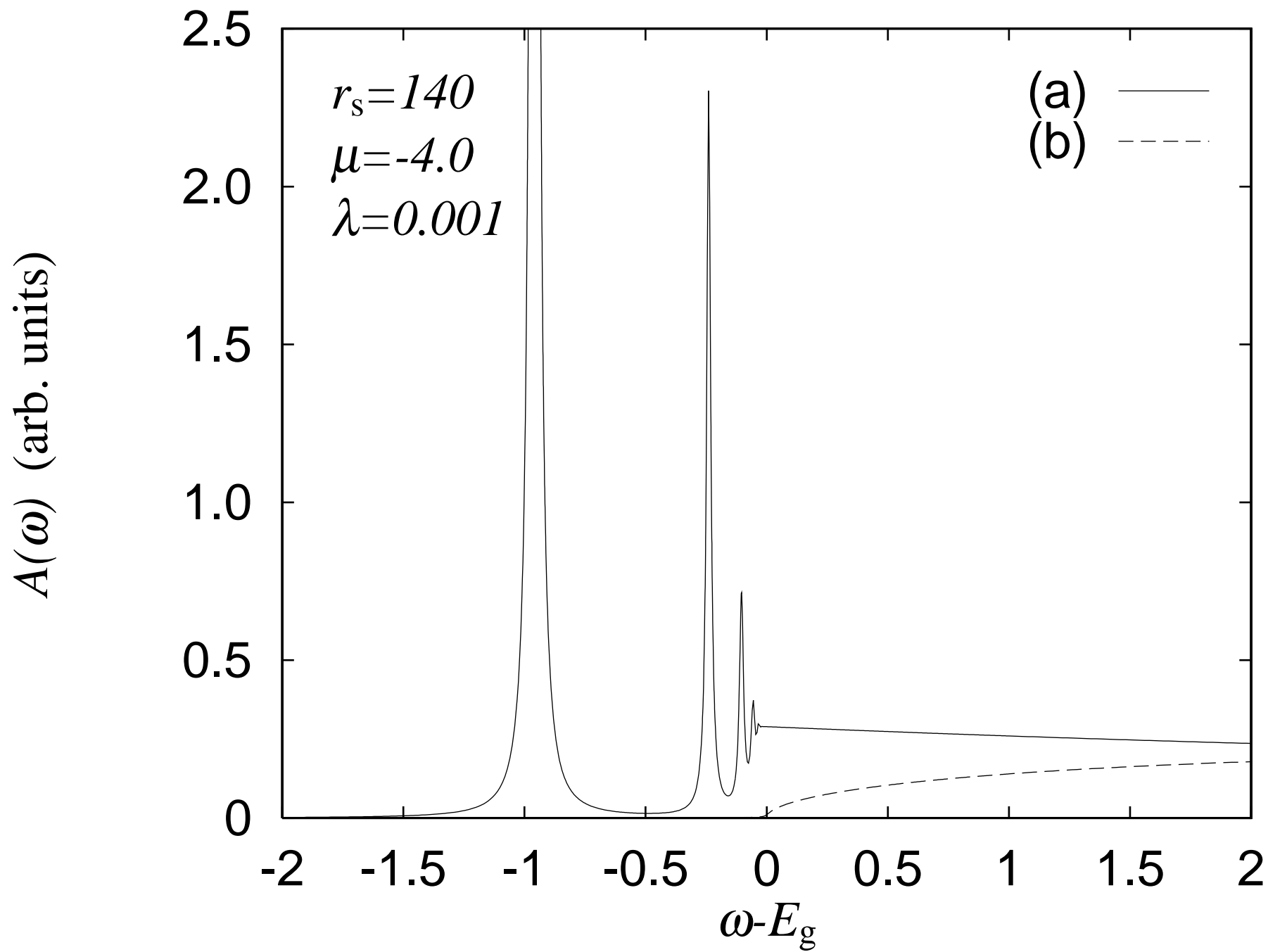
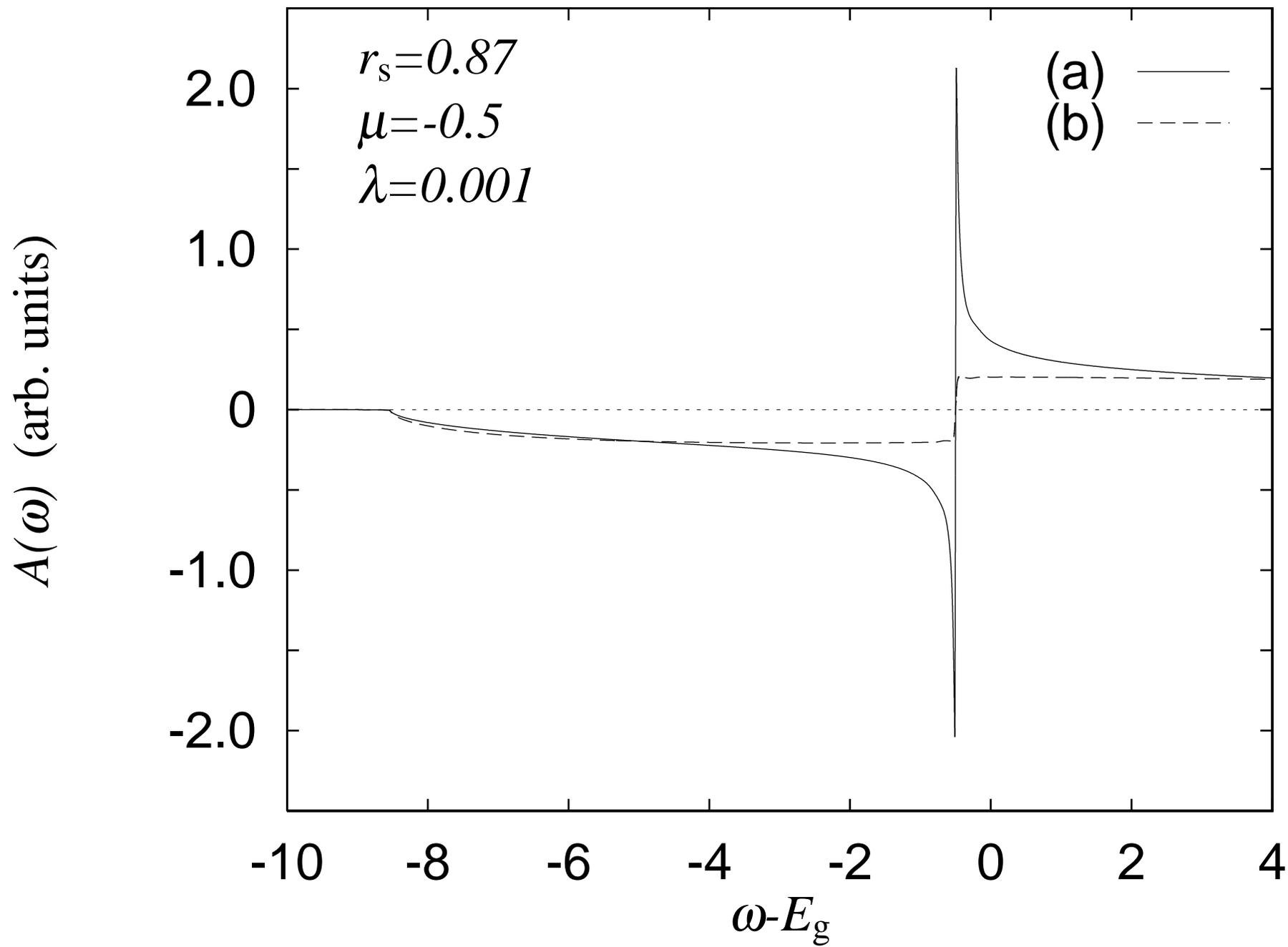
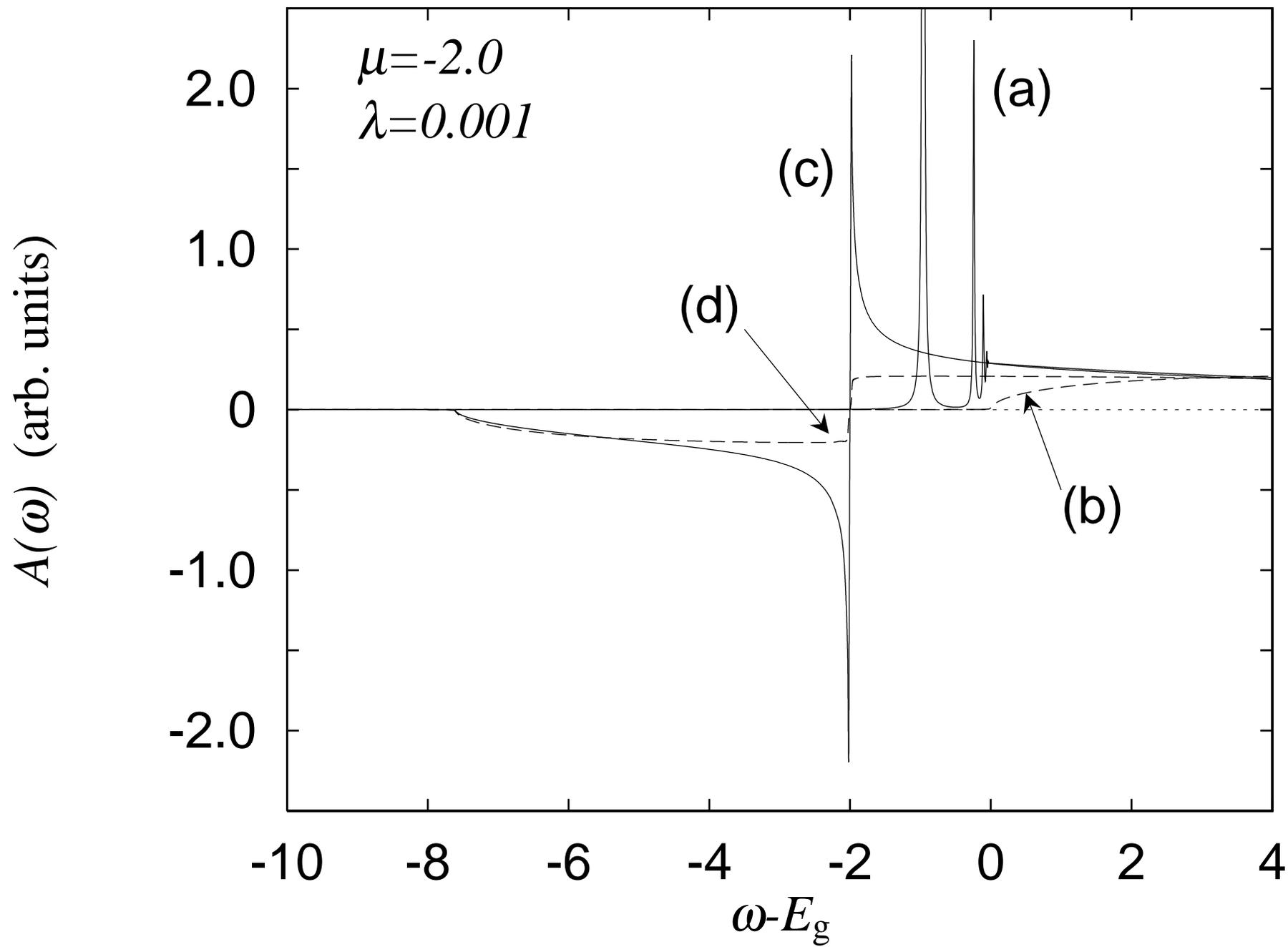
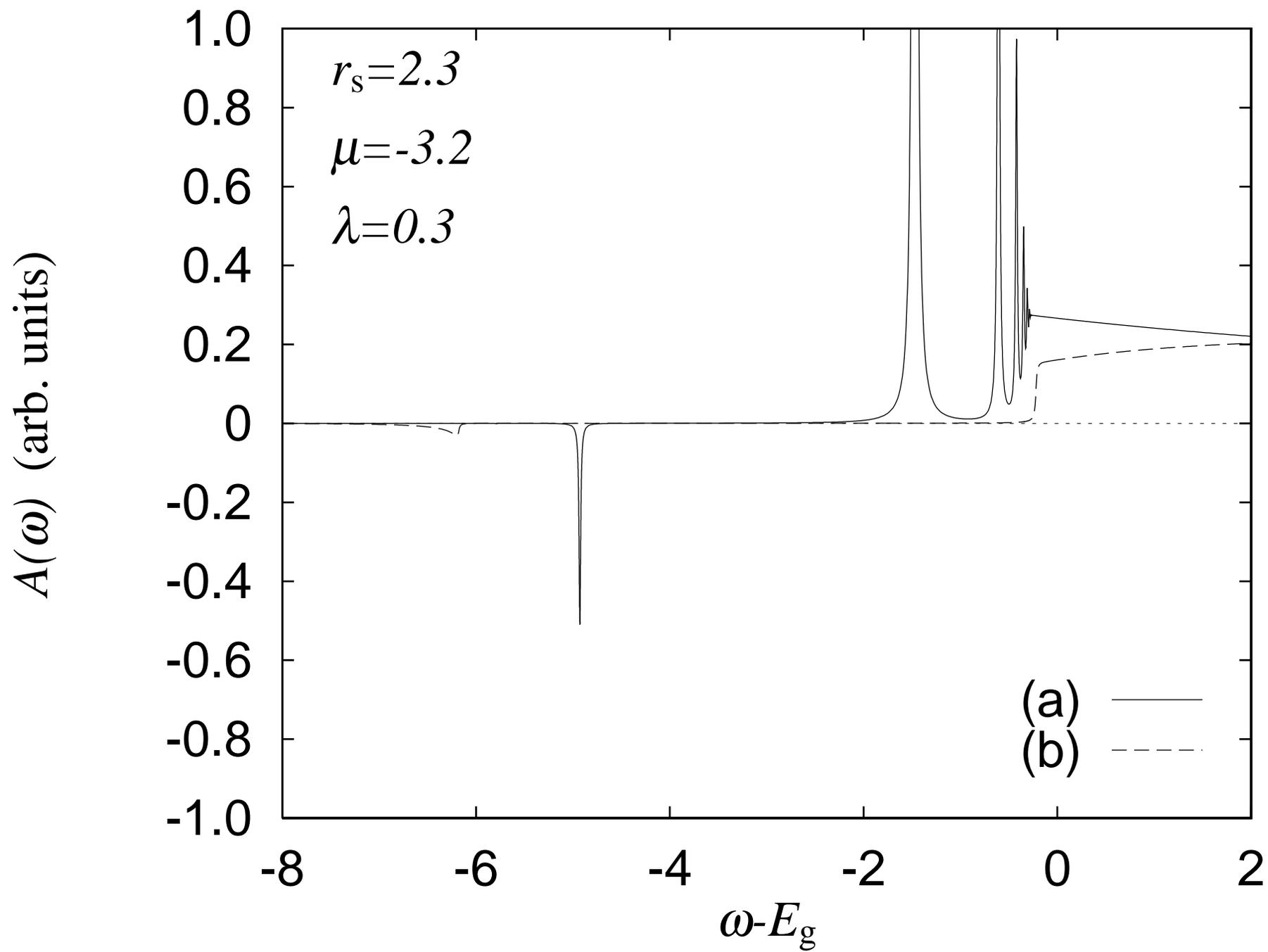
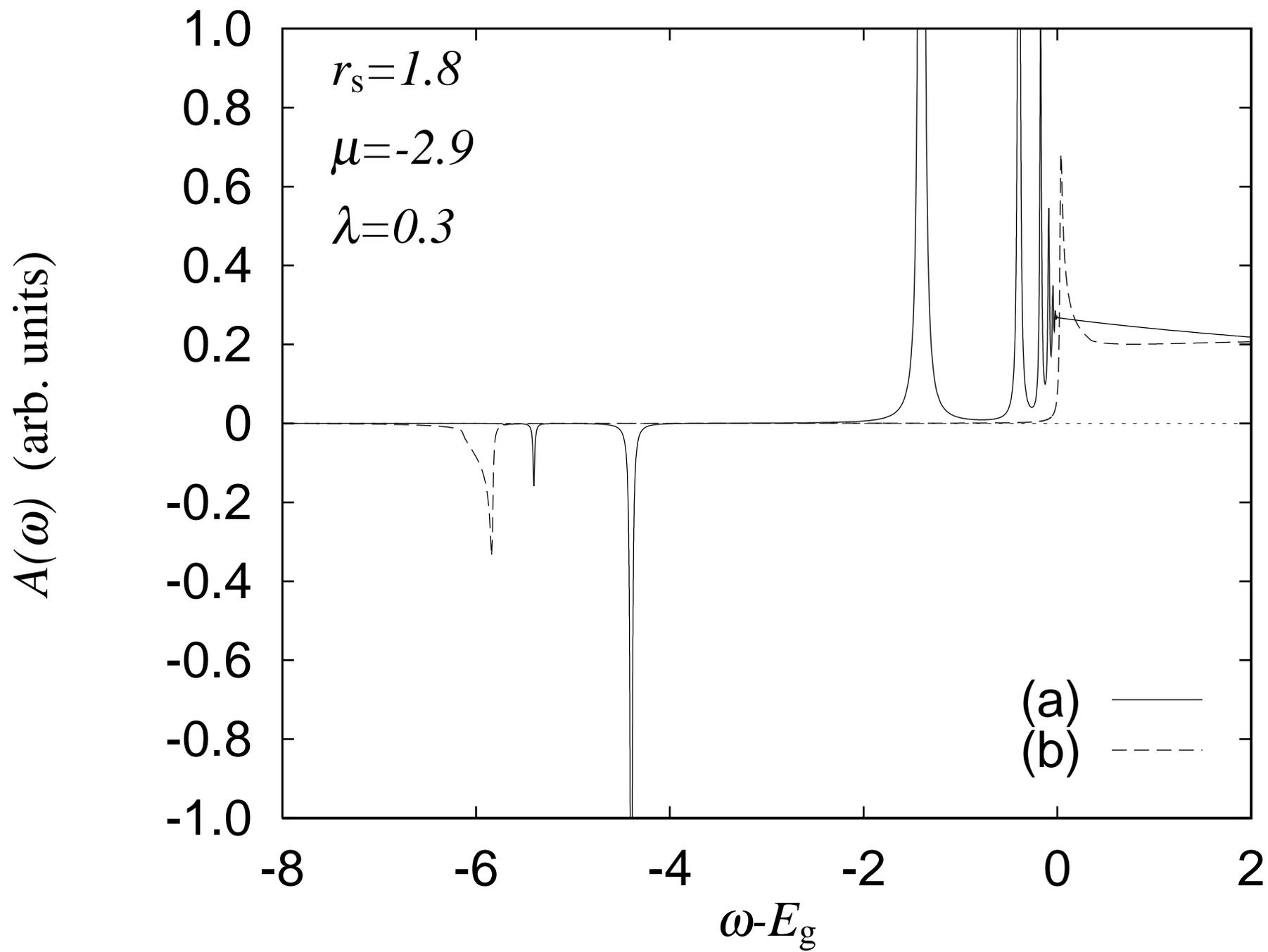
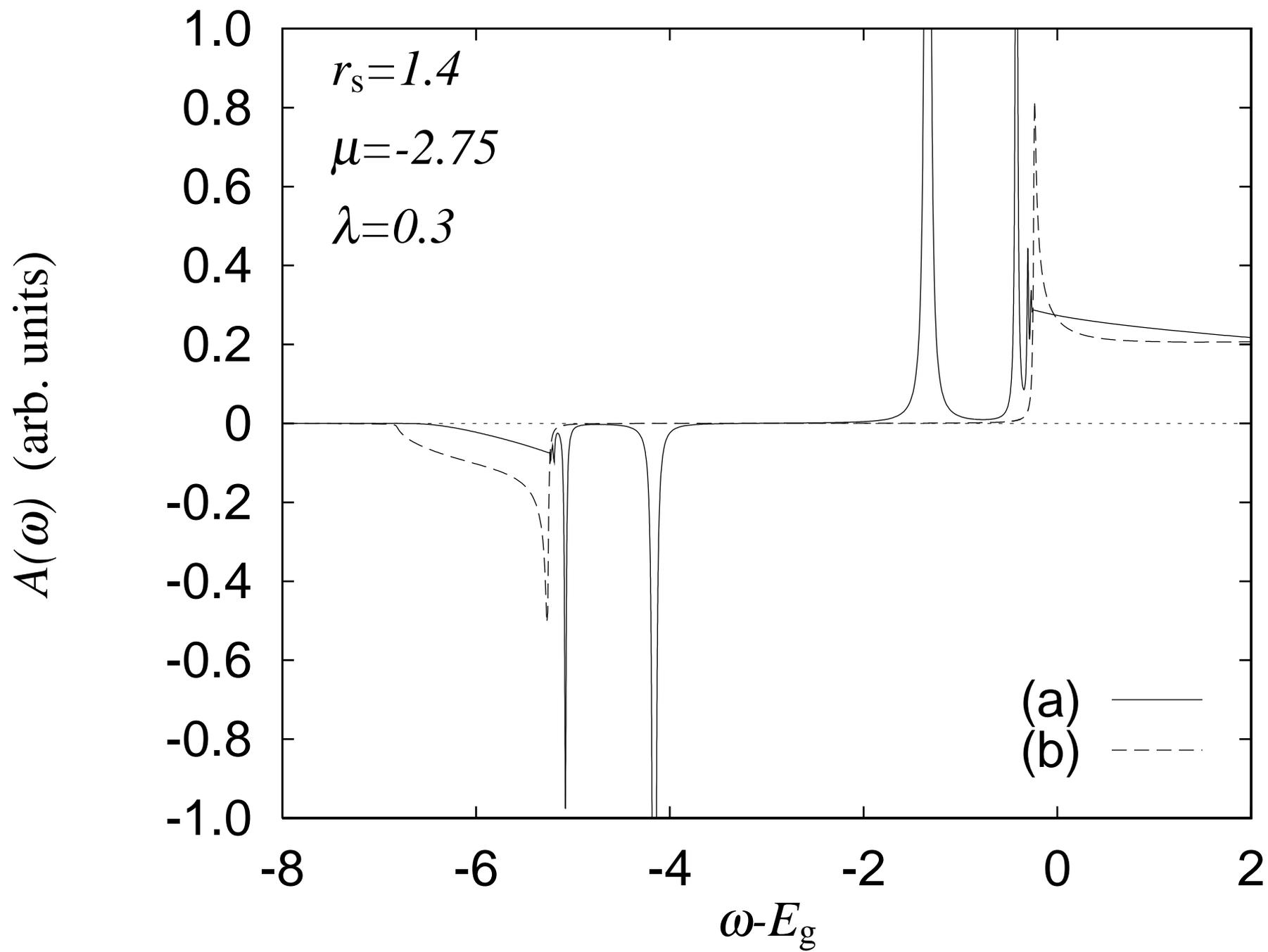
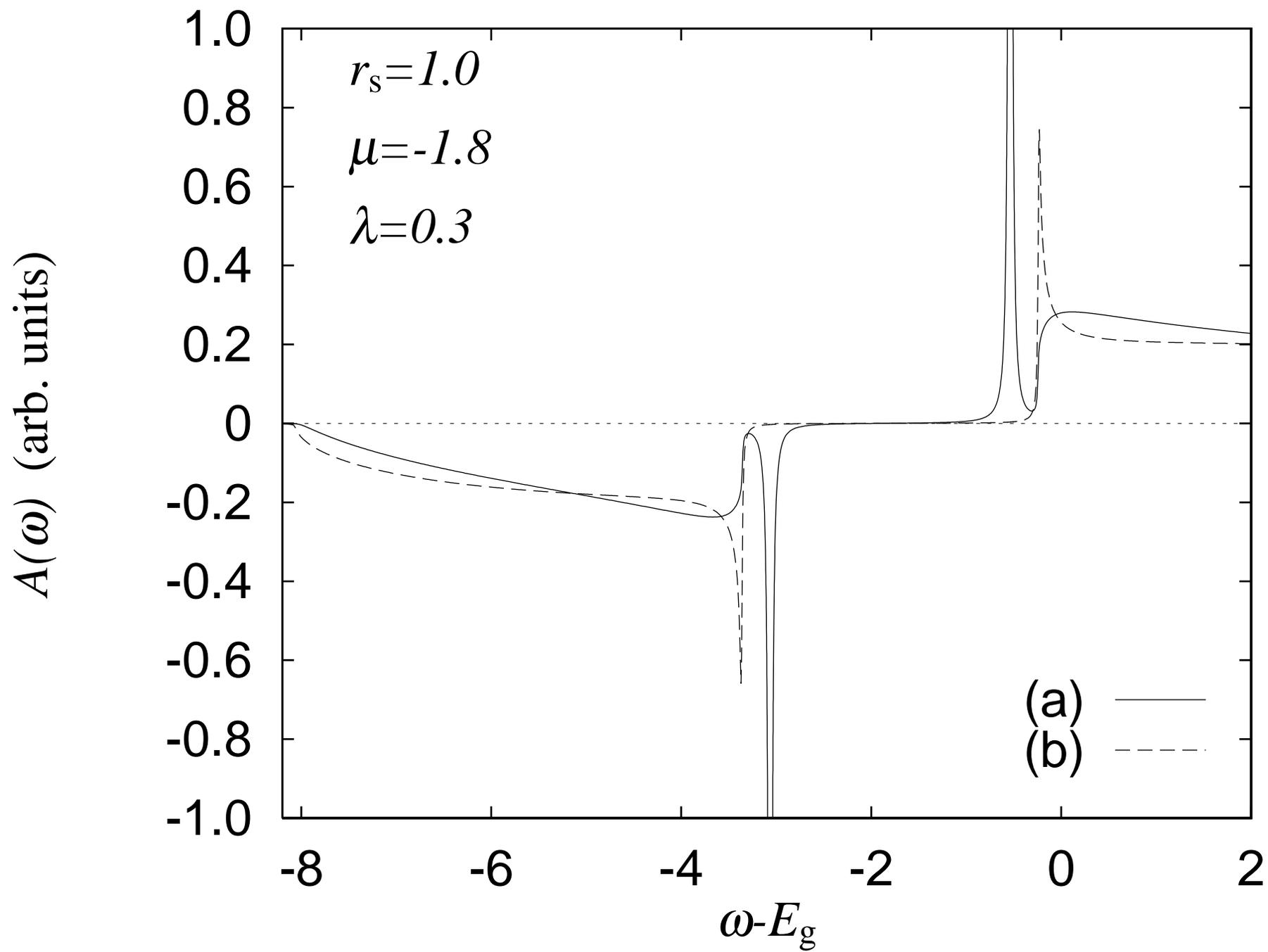
\end{document}